\documentclass[a4paper,11pt]{article}
\pdfoutput=1
\usepackage{jheppub}
\usepackage[T1]{fontenc}
\usepackage{graphicx}
\usepackage{amsmath}
\usepackage{amsmath}
\usepackage{xspace}
\usepackage{subfig}
\usepackage{xcolor}
\usepackage{mathtools}
\usepackage{braket}
\usepackage{multirow}

\title{Neutrinoless Double Beta Decay via  Light Neutralinos in R-Parity Violating Supersymmetry}

\affiliation[a]{\AddrUCL}
\affiliation[b]{\AddrSISSA}
\affiliation[c]{\AddrWash}

\author[a,b]{Patrick D. Bolton,} 
\emailAdd{patrick.bolton@ts.infn.it}
\author[a]{Frank F. Deppisch,} 
\emailAdd{f.deppisch@ucl.ac.uk}
\author[c]{P. S. Bhupal Dev}
\emailAdd{bdev@wustl.edu}

\newcommand{\AddrUCL}{Department of Physics and Astronomy, University College London,\\London WC1E 6BT, United Kingdom}

\newcommand{\AddrSISSA}{SISSA, International School for Advanced Studies, \\
INFN, Sezione di Trieste, Via Bonomea 265, I-34136 Trieste, Italy}

\newcommand{\AddrWash}{Department of Physics and McDonnell Center for the Space Sciences, Washington University, \\
St. Louis, MO 63130, U.S.A.}

\abstract{We perform a study of neutrinoless double beta ($0\nu\beta\beta$) decay mediated by the lightest neutralino of arbitrary mass in the minimal supersymmetric Standard Model (MSSM) under the presence of R-parity violating trilinear interactions. In this scenario, the exchange of the lightest neutralino can result in $0\nu\beta\beta$ decay of either long-range or short-range behaviour, depending on the neutralino mass. Using nuclear matrix elements calculated in the interacting boson model, we use an interpolation between the long- and short-range behaviours with an approximate formula. The non-observation of $0\nu\beta\beta$ decay is then used to place constraints on the supersymmetry parameter space, compatible with constraints from collider experiments. We compare these constraints to bounds from pion decays, CKM unitarity and big bang nucleosynthesis.}


\begin{document}	
\maketitle
\flushbottom

\section{Introduction}

In the minimal supersymmetric Standard Model (MSSM)~\cite{Fayet:1976et,Fayet:1977yc} it is possible to write down renormalisable and gauge-invariant terms in the superpotential $W$ which violate baryon ($B$) and lepton ($L$) number~\cite{Weinberg:1981wj},
\begin{align}
\label{eq:RPVsuperpotential}
    W \supset 
      \frac{1}{2}\lambda_{ijk} L_i L_j E_k^c + \lambda_{ijk}^{\prime} L_i Q_j D_k^c 
    + \frac{1}{2}\lambda_{ijk}^{\prime\prime} U_i^c D_j^c D_k^c + \kappa_i L_i H_u,
\end{align}
where $i,j,k$ are generation indices. Here, the $L$ and $Q$ chiral superfields contain the Standard Model (SM) lepton and quark $\text{SU}(2)_L$ doublets and their scalar (slepton and squark) partners. The $E^c$, $U^c$ and $D^c$ superfields respectively contain the SM charged lepton, up and down quark $\text{SU}(2)_L$ singlets and their scalar partners. Finally, the superfield $H_{u}$ contains the Higgs and Higgsino $\text{SU}(2)_L$ doublets.

The \textit{trilinear} terms with coefficients $\lambda_{ijk}$ and $\lambda'_{ijk}$ violate lepton number by one unit ($\Delta L = 1$), while the trilinear term with $\lambda''_{ijk}$ violates baryon number by one unit ($\Delta B = 1$). It can be shown that, at a given energy scale, the $\Delta L = 1$ \textit{bilinear} term $\kappa_{i}L_i H_u$ can be rotated away for complex-valued $\kappa_i$ upon a suitable redefinition of the lepton and Higgs superfields~\cite{Hall:1983id,Dreiner:2003hw,Chakraborty:2017dfg}. However, in presence of the trilinear terms, non-zero values of $\kappa_i$ can always be generated at other scales via renormalisation group running~\cite{deCarlos:1996ecd,Nardi:1996iy, Allanach:2003eb}. On the other hand, the bilinear coefficients are typically required to be small due to the $\lesssim$~eV neutrino masses~\cite{Hirsch:1998kc,Hirsch:2000ef,Mira:2000gg}. The phenomenology of trilinear interactions is therefore not affected by the presence of bilinear terms, except for specific phenomena such as neutrino oscillations. The trilinear coefficients are nevertheless subject to stringent limits from the non-observation of lepton and baryon number violating processes such as neutrinoless double beta ($0\nu\beta\beta$) decay, $(A, Z) \rightarrow(A, Z+2)+2 e^{-}$, and proton decay, $p\rightarrow \pi^0e^{+}, K^+\bar{\nu}$, respectively~\cite{Barbier:2004ez}. If a symmetry does not forbid such terms in the MSSM superpotential, this would imply a fine-tuning of the coefficients $\lambda_{ijk}$, $\lambda'_{ijk}$ and $\lambda''_{ijk}$.

It is therefore commonly assumed that the MSSM superpotential is invariant under $R$-parity~\cite{Farrar:1978xj}, defined as $R_{p} = (-1)^{3B+L+2S}$, where $S$ is the particle spin. The SM fields and their superpartners thus have $R_{p} = 1$ and $R_{p} = -1$ respectively. As a result, the terms in Eq.~\eqref{eq:RPVsuperpotential} are forbidden and the lightest supersymmetric particle (LSP) cannot decay to SM particles, and can thus be a dark matter candidate. However, there is no rigorous theoretical argument for $R$-parity conservation, and in fact, it has been argued that it might be more natural to include the $R$-parity violating (RPV) couplings rather than imposing $R$-parity by hand~\cite{Brust:2011tb}. One can instead impose the $Z_3$ baryon triality; this forbids the baryon number violating term but permits those that violate lepton number~\cite{Ibanez:1991pr, Dreiner:2007uj}. These lepton number violating terms have gained recent phenomenological interest in view of their ability to address current flavor anomalies in the muon anomalous magnetic moment and in semileptonic $B$-meson decays~\cite{Deshpande:2012rr, Biswas:2014gga, Zhu:2016xdg, Deshpande:2016yrv, Altmannshofer:2017poe, Das:2017kfo, Earl:2018snx, Trifinopoulos:2018rna, Wang:2019trs, Hu:2020yvs, Altmannshofer:2020axr, Dev:2021ipu, Bardhan:2021adp}. In this paper we will focus on the $0\nu\beta\beta$ constraints on and future prospects of the trilinear $\lambda'_{ijk}$ interactions. 

The contributions of the RPV terms with coefficients $\lambda_{i j k}^{\prime}$ and $\kappa_{i}$ to $0\nu\beta\beta$ decay have been studied before in the literature~\cite{Mohapatra:1986su,Vergados:1986td, Hirsch:1995ek, Faessler:1998qv, Hirsch:1998kc, Hirsch:2000jt,Allanach:2009xx, Li:2021fvw}; see Refs.~\cite{Rodejohann:2011mu, Deppisch:2012nb} for a review. As shown in Fig.~\ref{fig:RPV_SUSY}, $0\nu\beta\beta$ decay can proceed via the $\lambda'_{111}$ interaction at each vertex and the exchange of either a \textit{neutralino}~$\tilde{\chi}^0_i$ or \textit{gluino}~$\tilde{g}$. The bilinear coefficient $\kappa_1$ instead induces a mixing between the light neutrinos and neutralinos, and mass eigenstate neutralinos can be exchanged via SM charged current interactions. All previous studies have for simplicity considered the neutralinos and gluinos to be much heavier than the average momentum exchange of $0\nu\beta\beta$ decay, $m_{\tilde{\chi}^0_1}\gg p_{\text{F}}\sim 100~\text{MeV}$. However, as shown in Ref.~\cite{Dreiner:2009ic}, in the general MSSM with non-universal gaugino masses there
exists no model-independent laboratory bound on the mass of the lightest neutralino and an essentially massless neutralino is allowed by the experimental and observational data (see also Refs.~\cite{Hooper:2002nq, Bottino:2003iu, Belanger:2003wb, Lee:2007ai}).\footnote{In some versions of MSSM such as cMSSM, pMSSM and NMSSM, there does exist a lower bound on the lightest neutralino mass for it to be a good thermal dark matter candidate~\cite{Boehm:2013qva, Han:2014nba, Bramante:2015una, GAMBIT:2017zdo, Bagnaschi:2017tru,Barman:2020ylm}.} Then, similar to the exchange of sterile neutrinos~\cite{Bamert:1994qh, Benes:2005hn, Mitra:2011qr, Bolton:2019pcu}, neutralino exchange in $0\nu\beta\beta$ decay can either display long-range (if $m_{\tilde{\chi}^0_1}\ll p_{\text{F}}$) or short-range ($m_{\tilde{\chi}^0_1}\gg p_{\text{F}}$) behaviour. Examining this scaling behaviour and the resulting constraints on the RPV coupling $\lambda_{111}'$ for an arbitrary mass neutralino exchange is the main focus of this paper.

It is well-known that neutralinos also contribute at one-loop to the anomalous magnetic moment of the muon, $a_\mu \equiv (g-2)_\mu/2$~\cite{Grifols:1982vx,Moroi:1995yh, Carena:1996qa,Martin:2001st}. The persistence of the discrepancy between the theoretical and observed muon anomalous magnetic moment, first measured at BNL~\cite{Muong-2:2006rrc} and recently by the Muon g-2 experiment at Fermilab~\cite{Muong-2:2021ojo}, has led a number of papers to examine the implications for the MSSM parameter space~\cite{Gu:2021mjd,Chakraborti:2021dli,Han:2021ify,Baum:2021qzx,Baer:2021aax,Altmannshofer:2021hfu,Aboubrahim:2021ily}. Assuming that the selectron and smuon masses are degenerate, this work will compare the favoured region from the $(g-2)_\mu$ discrepancy to the corresponding excluded regions from the non-observation of $0\nu\beta\beta$ decay.

The rest of the paper is organised as follows: In Section~\ref{sec:MSSMconventions} we will first review our conventions for the neutralino mass matrix and highlight the limit in which the lightest neutralino is (almost) massless. We also define the (RPV) MSSM couplings relevant for $0\nu\beta\beta$ decay and $(g-2)_\mu$. In Section~\ref{sec:RPVconstraints}, we outline the existing experimental constraints on the MSSM parameters relevant to $0\nu\beta\beta$ decay. These are the RPV coupling $\lambda_{111}'$, the lightest neutralino mass $m_{\tilde{\chi}^0_1}$, the selectron and up/down squark masses $m_{\tilde{e}_L}$, $m_{\tilde{u}_L}$, $m_{\tilde{d}_R}$ and gluino mass $m_{\tilde{g}}$. In Section~\ref{sec:RPV0vbb} we review the contribution of neutralino and gluino exchange to the $0\nu\beta\beta$ decay rate and the corresponding half-life $T^{0\nu}_{1/2}$. Making use of an interpolating formula, we generalise to the case where the neutralino can be either lighter or heavier than the average momentum exchange of $0\nu\beta\beta$ decay. In Section~\ref{sec:RPVexcl}, we derive constraints on the RPV coupling $\lambda_{111}'$ as a function of the neutralino mass $m_{\tilde{\chi}^0_1}$, requiring that all collider limits on the other sparticle masses apply. We also plot the excluded regions in the $(m_{\tilde{e}_L},m_{\tilde{\chi}_1^0})$ and $(\lambda_{111}',m_{\tilde{e}_L})$ planes. In these parameter spaces we compare the $0\nu\beta\beta$ decay excluded regions to existing experiments limits and naive bounds derived from big bang nucleosynthesis (BBN). We also contrast these excluded regions to the favoured region from the $(g-2)_\mu$ discrepancy. We finally conclude in Section~\ref{sec:ch6-concl}. The analytic expressions for the neutralino partial and total decay widths are given in Appendix~\ref{sec:decay_rates}, and the relevant details for the $(g-2)_\mu$ calculation are relegated to Appendix~\ref{sec:g-2}.  

\section{MSSM Conventions}
\label{sec:MSSMconventions}

In the MSSM, the $4\times 4$ mass matrix for the four neutral gauginos (the bino $\tilde{B}$, neutral wino $\tilde{W}$ and two Higgsinos $\tilde{H}_d^0$, $\tilde{H}_u^0$) is written as~\cite{Nilles:1983ge,Haber:1984rc,Choi:2001ww}
\begin{align}
\label{eq:gauginomass}
    \mathcal{M}_{\tilde{\chi}}=
    \begin{pmatrix}
        M_1              & 0                & -m_Z s_W c_\beta &  m_Z s_W s_\beta \\
        0                & M_2              &  m_Z c_W c_\beta & -m_Z c_W s_\beta \\
        -m_Z s_W c_\beta &  m_Z c_W c_\beta & 0                & -\mu             \\
         m_Z s_W s_\beta & -m_Z c_W s_\beta & -\mu             & 0
\end{pmatrix},
\end{align}
with $c_W = \cos\theta_W$, $s_W = \sin\theta_W$, $c_\beta = \cos\beta$ and $s_\beta = \sin\beta$. Here, $\theta_W$ is the SM weak mixing angle, $m_Z$ the $Z$ boson mass, $M_1$ and $M_2$ are the bino and neutral wino masses, $\mu$ is the Higgsino mass parameter and $\tan\beta = v_u/v_d$ is the ratio of the vacuum expectation values of the two Higgs fields $H_u$ and $H_d$.

The mass matrix $\mathcal{M}_{\tilde{\chi}}$ can be diagonalised by a unitary matrix as $N^T\mathcal{M}_{\tilde{\chi}}N$, giving four mass eigenstate neutralinos,
\begin{align}
\tilde{\chi}_{i}^{0}=N_{i 1} \tilde{B}+N_{i 2} \tilde{W}^{0}+N_{i 3} \tilde{H}_{d}^{0}+N_{i 4} \tilde{H}_{u}^{0}\,.
\end{align}
It is commonly assumed that the experimental lower bound on the lightest chargino mass, $m_{{\tilde{\chi}}^{\pm}_1}> 94$~GeV~\cite{DELPHI:2003uqw}, sets a lower bound on the parameters $M_2,|\mu|\gtrsim 100$~GeV~\cite{Barger:2005hb}. Using the relation $M_1 = \frac{5}{3}\tan^2\theta_W M_2$ from grand unified theories, this can then be used to set a lower bound on $M_1$ and hence the lightest neutralino mass $m_{\tilde{\chi}^0_1}$. The experimental lower bound $m_{\tilde{\chi}^0_1}>46$~GeV set by DELPHI furthermore assumes $M_2<1$~TeV, $|\mu|< 2$~TeV, $\tan\beta>5$ and that $\tilde{\chi}^0_1$ is the LSP~\cite{DELPHI:2003uqw}.

If $M_1$, $M_2$ and $\mu$ are instead chosen to be free parameters, it is in fact possible for the lightest neutralino to be essentially massless~\cite{Dreiner:2009ic}. To see this, one can set the determinant of the mass matrix $\mathcal{M}_\mathcal{\tilde{\chi}}$ to zero and rearrange for $M_1$,
\begin{align}
\mathrm{det}\left(\mathcal{M}_{\tilde{\chi}}\right)=0 \quad \Rightarrow \quad M_{1}=\frac{m_{Z}^{2} M_{2} s_W^2 s_{2\beta}}{\mu M_{2}-m_{Z}^{2} c_W^2s_{2\beta}}\approx \frac{2m_{Z}^2s_W^2}{\mu t_\beta}\,,
\end{align}
where $s_{2\beta} = \sin 2\beta$ and $t_\beta=\tan\beta$. The approximate relation neglects the second term in the denominator and uses $s_{2\beta} = 2t_\beta/(1+t^2_\beta)\approx 2/t_\beta$ for $t_\beta\gtrsim 3$. For fixed values of $M_2$, $\mu$ and $t_\beta$, $M_1$ can always take values such that the lightest neutralino $\tilde{\chi}_1^0$ is massless. In this limit, the lightest neutralino is predominantly bino-like and couples to fermions and sfermions via the weak hypercharge, $Y=Q^f-I^f_3$, where $Q^f$ and $I^f_3$ are the electric charge and the third component of the weak isospin of the fermion and sfermion, respectively.

In general, the neutralino-fermion-sfermion interaction is given by~\cite{Haber:1984rc}
\begin{align}
\label{eq:LRint}
\mathcal{L}_{\tilde{\chi} f \tilde{f}} \supset \sqrt{2}g \bar{f}_{j}  \big[(V^L_{f_j\tilde{\chi}^0_{i}})_{j\kappa} P_R + (V^R_{f_j\tilde{\chi}^0_{i}})_{j\kappa} P_L\big]  \tilde{\chi}^0_{i}\tilde{f}_{\kappa} +\mathrm{h.c.}\,,
\end{align}
where $g$ is the $\mathrm{SU}(2)_L$ coupling constant, $P_{L,R}=(1\mp \gamma^5)/2$ are the chirality projection operators, and the indices $i$, $j$ and $\kappa$ indicate the neutralino mass eigenstate, fermion generation and sfermion mass eigenstate respectively. The left and right-handed couplings are given, respectively, by
\begin{align}
\label{eq:LRcouplings}
(V^L_{f_j\tilde{\chi}^0_{i}})_{j\kappa} &= V^{LL}_{f_j\tilde{\chi}^0_{i}} (U^{L}_{\tilde{f}})_{j\kappa}+ V^{LR}_{f_j\tilde{\chi}^0_{i}} (U^{R}_{\tilde{f}})_{j\kappa} \,, \\
(V^R_{f_j\tilde{\chi}^0_{i}})_{j\kappa} &= V^{RL}_{f_j\tilde{\chi}^0_{i}} (U^{L}_{\tilde{f}})_{j\kappa}+ V^{RR}_{f_j\tilde{\chi}^0_{i}} (U^{R}_{\tilde{f}})_{j\kappa} \,,
\label{eq:LRcouplings2}
\end{align}
where
\begin{align}
\label{eq:VLL}
  \begin{split}
    V^{LL}_{f_j\tilde{\chi}^0_{i}} &= -\big(Q^f-I^f_{3}\big)t_{W}N_{i 1}-T_{3} N_{i 2}\,,\\
    V^{LR}_{u_j\tilde{\chi}^0_{i}} &= V^{RL}_{u_j\tilde{\chi}^0_{i}} = -\frac{m_{u_j}N_{i4}}{2m_W s_\beta}\,, 
  \end{split}
  \begin{split}
    V^{RR}_{f_j\tilde{\chi}^0_{i}}  &= Q^{f} t_{W} N_{i 1}\,,\\
    V^{LR}_{d_j\tilde{\chi}^0_{i}} &= V^{RL}_{d_j\tilde{\chi}^0_{i}} = -\frac{m_{d_j}N_{i3}}{2m_W c_\beta}\,,
  \end{split}
\end{align}
with $u_j$ indicating up-type quarks and neutrinos and $d_j$ down-type quarks and charged leptons. However, in the following we will assume for simplicity a completely bino-like lightest neutralino, so $N_{11}=1$ and $N_{12}=N_{13}=N_{14}=0$. In Eqs.~\eqref{eq:LRcouplings} and \eqref{eq:LRcouplings2}, $U_{\tilde{f}}^L$ and $U_{\tilde{f}}^R$ are $3\times 6$ matrices rotating the left- and right-handed sfermions to six mass eigenstate sfermions. If no mixing is present between generations, the non-zero elements of the matrices are
\begin{align}
&(U_{\tilde{f}}^L)_{jj} = \cos\theta_{\tilde{f}_j}\,, \quad (U_{\tilde{f}}^L)_{j(j+3)} = -\sin\theta_{\tilde{f}_j}\,, \nonumber \\
&(U_{\tilde{f}}^R)_{jj} = \sin\theta_{\tilde{f}_j}\,, \quad (U_{\tilde{f}}^R)_{j(j+3)} = \cos\theta_{\tilde{f}_j}\,,
\end{align}
where $\theta_{\tilde{f}_j}$ is the left-right mixing angle for the sfermion $\tilde{f}_j$.

The RPV interactions of relevance to $0\nu\beta\beta$ decay can be found by expanding the chiral superfields in the $\lambda'_{ijk}$ term in Eq.~\eqref{eq:RPVsuperpotential} as $L_iQ_jD_k^c\equiv \left(\epsilon_{ab}L_i^aQ_j^b\right)D_k^c$ (where $a,b$ are $\text{SU(2)}_L$ indices and $\epsilon_{ab}$ is the  antisymmetric tensor) in standard four-component Dirac notation:
\begin{align}
\label{eq:L_LQD}
\mathcal{L}_{L Q D} \supset \lambda_{ijk}^{\prime}\Big[&\tilde{e}^*_{L i} \bar{u}_{Lj} d_{Rk}+\tilde{u}^*_{L j} \bar{e}_{Li}  d_{Rk}+\tilde{d}_{R k} \bar{u}_{Lj} e^c_{Ri} \nonumber\\
&-\tilde{\nu}^*_{Li}\bar{d}_{Lj}d_{Rk}-\tilde{d}^*_{L j}\bar{\nu}_{Li} d_{Rk}-\tilde{d}_{Rk}\bar{d}_{Lj}\nu^c_{Ri}\Big]+\mathrm{h.c.}\,,
\end{align}
where the indices $i$, $j$ and $k$ indicate the (s)fermion generation, and the superscript $c$ denotes charge conjugation, e.g. $\nu^c_{Ri}=\left(\nu_i^c\right)_R$ is the charge conjugate of $\nu_{Li}$. We will see that the first three terms in Eq.~\eqref{eq:L_LQD} along with the interactions in Eq.~\eqref{eq:LRint} contribute to $0\nu\beta\beta$ decay at tree-level. The combination of these interactions also allows neutralinos to decay via channels such as $\tilde\chi^0_i\rightarrow \nu \pi^0$, $e^- \pi^+$, $\nu \gamma$~\cite{Baltz:1997gd,Kim:1998mu}. We give the decay rates for these processes in Appendix~\ref{sec:decay_rates}.

\section{Experimental Constraints}
\label{sec:RPVconstraints}

To date, no experimental evidence of supersymmetry has been established~\cite{Zyla:2020zbs}. Null results therefore translate to excluded regions in the MSSM parameter space. The most general MSSM superpotential, however, depends on 124 free parameters (19 SM parameters and 105 additional parameters), making a phenomenological analysis impractical. The number of free parameters is usually reduced by assuming a particular supersymmetry-breaking scheme~\cite{Intriligator:2007cp}, such as gravity- or gauge-mediated mechanisms. The gravity-mediated case can be simplified down even further to the so-called \textit{constrained} MSSM (cMSSM), with only 5 additional free parameters~\cite{Kane:1993td}. The null results from ATLAS and CMS have now excluded the cMSSM for TeV-scale superpartner masses at 90\% CL~\cite{Bechtle:2015nua}.

A less predictive, and necessarily more poorly constrained model, is the \textit{phenomenological} MSSM (pMSSM), with 19 additional free parameters~\cite{MSSMWorkingGroup:1998fiq}. It requires that supersymmetry is not a source of $CP$ violation and flavor-changing neutral currents, and that the first two generations of sfermions are degenerate. ATLAS and CMS have interpreted data from Run 1 of the LHC in the pMSSM, determining the fraction of models that are excluded in a given region of the parameter space~\cite{Aad:2015baa,Khachatryan:2016nvf}. We show to the left of Table~\ref{tab:SUSYlimits} the rough values of the relevant selectron, up and down squark and gluino masses ($m_{\tilde{e}_L}$, $m_{\tilde{u}_L}$, $m_{\tilde{d}_R}$ and $m_{\tilde{g}}$) below which 90\% of models are excluded. We take these to be approximate lower bounds on these sparticle masses, independent of the lightest neutralino mass $m_{\tilde{\chi}^0_1}$.

ATLAS and CMS have also conducted searches for slepton, squark and gluino production and subsequent decays to SM particles (either leptonic or jet final states) and neutralinos, which are undetected and contribute to missing transverse energy. Derived lower bounds on the masses $m_{\tilde{e}_L}$, $m_{\tilde{u}_L}$, $m_{\tilde{d}_R}$ and $m_{\tilde{g}}$ depend on the neutralino mass $m_{\tilde{\chi}^0_1}$ in the region $m_{\tilde{\chi}^0_1} < m_{\tilde{X}}$ (where the decay process is kinematically allowed). 

To the right of Table~\ref{tab:SUSYlimits} we give the lower bounds on $m_{\tilde{e}_L}$, $m_{\tilde{u}_L}$, $m_{\tilde{d}_R}$ and $m_{\tilde{g}}$ for a massless neutralino and the indicated decay process. The branching ratios for these decays are assumed to be 100\%, requiring the lightest neutralino and chargino to be purely bino and wino, respectively. Generically, this can be achieved for large values of $\mu$ and $M_2$. Of course, in the presence of the RPV coupling $\lambda_{111}'$, additional decay modes are open for the sleptons. We make the assumption that the decays $\tilde{e}_L\rightarrow e \tilde{\chi}^0_1$, $\tilde{q}\rightarrow q \tilde{\chi}^0_1$ and $\tilde{g}\rightarrow q\bar{q} \tilde{\chi}^0_1$ still dominate even for $\lambda_{111}'\neq 0$, which is justified if $\lambda_{111}'$ is small. The bounds weaken considerably for larger $m_{\tilde{\chi}^0_1}$, and often a different decay mode provides a more stringent lower bound. 

For selectrons, the ALEPH experiment imposes a lower bound of $m_{\tilde{e}_L} > 107$~GeV for \textit{any} value of the neutralino mass below $m_{\tilde{e}_L}$~\cite{Heister:2002jca}. It should be noted that these analyses rely on the assumption that the neutralino does not decay within the detector. If a large RPV coupling $\lambda_{111}'$ exists, we will see that a LSP or NLSP (next-to-lightest supersymmetric particle) neutralino can decay. For these limits to apply, $\lambda_{111}'$ must be small enough for the neutralino to be long-lived. However, a long-lived neutralino can itself provide distinct displaced vertex signatures at future beam-dump experiments and downstream detectors at collider experiments~\cite{Gorbunov:2015mba,deVries:2015mfw,Helo:2018qej,Dercks:2018eua,Dercks:2018wum,Wang:2019orr,Wang:2019xvx,Dreiner:2020qbi,Dey:2020juy}. Limits on $\lambda_{111}'$ (assuming non-zero $\lambda_{112}'$) have recently been found by considering long-lived neutralinos, produced in cosmic ray air showers, decaying inside the Super-Kamiokande detector~\cite{Candia:2021bsl}. The OPAL experiment in contrast searched for direct RPV decays, setting a lower bound of $m_{\tilde{e}_L} > 87$~GeV for $m_{\tilde{\chi}^0_1} > m_{\tilde{e}_L}$~\cite{Abbiendi:2003rn}.

\begin{table}[]
	\centering
	\renewcommand{\arraystretch}{1.25}
	\setlength\tabcolsep{5.2pt}
	\begin{tabular}{ccccc}
		\hline
		\multirow{2}{*}{$\tilde{X}$} & \multicolumn{4}{c}{$m_{\tilde{X}}$ Lower Bound $[\mathrm{GeV}]$} \\ \cline{2-2} \cline{4-5}
		&  pMSSM~\cite{Aad:2015baa}  & & $m_{\tilde{\chi}^0_{1}}=0$ & Experiment \\ \hline
		$\tilde{e}_L$ & $\sim 90$ &   &  $700$ ($700$) & $\tilde{e}_L\rightarrow e\tilde{\chi}^0_1$, ATLAS~\cite{Aad:2019vnb} (CMS~\cite{Sirunyan:2020eab}) \\
		$\tilde{u}_L$, $\tilde{d}_R$  & $\sim600$ &  & $1900$ ($1750$) & $\tilde{q}\rightarrow q\tilde{\chi}_1^{0}$, ATLAS~\cite{ATLAS-CONF-2019-040} (CMS~\cite{Sirunyan:2019xwh}) \\
		$\tilde{g}$ & $\sim1200$ & &  $2350$ ($2000$)& $\tilde{g}\rightarrow q\bar{q}\tilde{\chi}_1^{0}$, ATLAS~\cite{ATLAS-CONF-2019-040} (CMS~\cite{Sirunyan:2019ctn})\\ \hline
		$\tilde{\psi}$ & \multicolumn{3}{c}{$1.35\times 10^{-14}$}& $e^{+}e^{-}\rightarrow \tilde{\psi}\tilde{\psi}\gamma$, L3~\cite{Achard:2003tx}\\ \hline
	\end{tabular}
	\caption{Lower limits on the masses of the superpartners relevant to $0\nu\beta\beta$ decay when an RPV coupling $\lambda_{111}'$ is present, approximately in the pMSSM by excluding 90\% of models, and from searches for superpartners decaying to a massless (purely bino) neutralino with 100\% branching ratio at ATLAS and CMS. Also shown is the lower bound on the gravitino $\tilde\psi$ mass from L3.}
	\label{tab:SUSYlimits}
\end{table}

The presence of the RPV coupling $\lambda_{111}'$ and a down squark ($\tilde{d}_{R}$) also leads to additional semileptonic quark decays at tree-level, changing the experimentally measured CKM matrix element $|V_{ud}|_{\text{exp}}$ to
\begin{align}
\label{eq:VCKM_RPV}
\left|V_{u d}\right|_{\text{exp}}^{2}=|V_{u d}|^2\left|1+\frac{1}{4\pi\alpha_2}\frac{m_{W}^{2}}{m_{\tilde{d}_{R}}^{2}}\frac{|\lambda_{111}^{\prime}|^2}{V_{ud}}\right|^{2}\,,
\end{align}
where $|V_{ud}|$ is the value in the SM~\cite{Barger:1989rk}. Taking the current best-fit value $\left|V_{u d}\right|_{\text{exp}}^{2} = 0.97420 \pm 0.00021$~\cite{Zyla:2020zbs}, we set $\left|V_{u d}\right|_{\text{exp}}^{2}$ to the maximally allowed value and $\left|V_{u d}\right|^{2}$ to the best-fit value to obtain an upper limit on  
\begin{align}
|\lambda_{111}'| \leq 0.012\left(\frac{m_{\tilde{d}_R}}{100~\mathrm{GeV}}\right) \, .
\label{eq:lamCKM}
\end{align}
A non-zero $\lambda'_{111}$ and down squark also contribute at tree-level to pion decays, affecting the ratio of decays to the first and second generation as
\begin{align}
\label{eq:Rpi_RPV}
R^{(\pi)}_{e/\mu,\,\text{exp}}=\frac{\Gamma\left(\pi^{-} \rightarrow e^{-} \bar{\nu}_{e}\right)}{\Gamma\left(\pi^{-} \rightarrow \mu^{-} \bar{\nu}_{\mu}\right)}\simeq R^{(\pi)}_{e/\mu}\left[1+\frac{1}{2\pi\alpha_2}\frac{m_{W}^{2}}{m_{\tilde{d}_{R}}^{2}}\frac{|\lambda_{111}^{\prime}|^2}{V_{u d}}\right]\,.
\end{align}
We insert into this expression the theoretical prediction~\cite{Cirigliano:2007xi,Bryman:2011zz} and measured value~\cite{Aguilar-Arevalo:2015cdf} of this ratio,
\begin{align}
R^{(\pi)}_{e/\mu} &= (1.2352\pm 0.0001)\times10^{-4}\,,\\
R^{(\pi)}_{e/\mu,\,\text{exp}} &= (1.2344 \pm 0.0023 \pm 0.0019) \times 10^{-4}\,,
\end{align}
respectively. Rearranging for $\lambda_{111}'$ in Eq.~\eqref{eq:Rpi_RPV} thus gives an upper bound
\begin{align}
|\lambda_{111}'| \leq 0.014\left(\frac{m_{\tilde{d}_R}}{100~\mathrm{GeV}}\right) \,, 
\label{eq:lamRpi}
\end{align}
which is slightly weaker than that obtained from $V_{ud}$ in Eq.~\eqref{eq:lamCKM}.

As explored in Ref.~\cite{Kim:1998mu}, a neutralino that can decay via RPV interactions can also impact the formation of light elements during BBN. To avoid such interference, we naively require the neutralino lifetime $\tau_{\tilde{\chi}^0_i} = \Gamma^{-1}_{\tilde{\chi}^0_i}$ be shorter than 1 second, as has been considered before in the context of sterile neutrinos, for instance~\cite{Gorbunov:2007ak, Ruchayskiy:2012si, Sabti:2020yrt, Boyarsky:2020dzc}. The rates of RPV neutralino decay modes are tabulated in Appendix~\ref{sec:decay_rates} and are summed to compute the total neutralino decay width $\Gamma_{\tilde{\chi}^0_i}$. The condition $\tau_{\tilde{\chi}^0_i} < 1 ~\mathrm{s}$ then translates to an excluded region in the space of MSSM parameters entering $\tau_{\tilde{\chi}^0_i}$. For example, for fixed $m_{\tilde{u}_L}$, $m_{\tilde{d}_R}$ and $m_{\tilde{g}}$, we obtain an excluded region in the ($\lambda_{111}',m_{\tilde{e}_L}$) parameter space. We will see that this constraint heavily depends on the mass of the gravitino $m_{\tilde{\psi}}$. A lower bound on the gravitino mass, $m_{\tilde{\psi}} > 1.35\times 10^{-14}$~GeV, has been set by the L3 experiment~\cite{Achard:2003tx}. The gravitino can therefore be the LSP and the lightest neutralino the NLSP for the BBN constraint to apply. As mentioned previously, a gravitino or neutralino LSP could be a promising dark matter candidate~\cite{Steffen:2006hw,Buchmuller:2007ui,Covi:2009bk,Jean-Louis:2009zrs,Buchmuller:2009fm,Benakli:2017whb,Drees:2018dsj,Gu:2020ozv}. As we consider a wide range of possible masses for the gravitino, lightest neutralino and other SUSY partners in this work, for simplicity we do not attempt to incorporate a viable component of dark matter.

\section{R-Parity Violating Contributions to \texorpdfstring{$0\nu\beta\beta$}{0vbb} Decay}
\label{sec:RPV0vbb}

Previous studies have considered the contribution of the RPV couplings $\lambda_{111}'$ and $\kappa_1$ to $0\nu\beta\beta$ decay for large superpartner masses, i.e. above the average momentum exchange of the process ($m_{\tilde{X}}\gg p_{\text{F}} \sim 100$~MeV)~\cite{Mohapatra:1986su,Vergados:1986td,Hirsch:1995ek,  Faessler:1998qv,Hirsch:1998kc, Hirsch:2000jt, Allanach:2009xx, Li:2021fvw}. In this limit the process can be taken to be a \textit{short-range} $0\nu\beta\beta$ decay mechanism, as opposed to a \textit{long-range} mechanism, like  the exchange of light Majorana neutrinos.

\begin{figure}[t]
	\centering
	\includegraphics[trim=0 -0.4cm 0 -0.4cm,width=4cm]{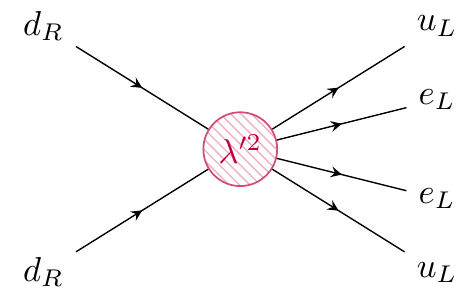}
	\put(5,40){\normalsize{$ = $}}
	\hspace{1em}
	\includegraphics[width=4cm]{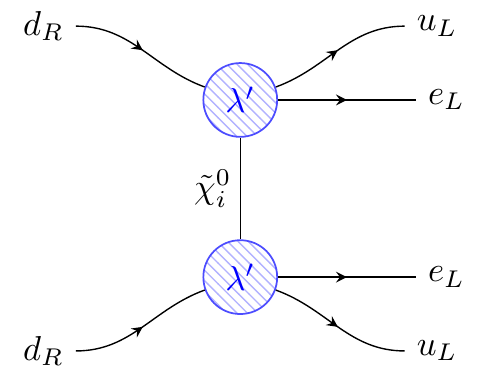}
	\put(5,40){\normalsize{$ + $}}
	\hspace{1em}
	\includegraphics[width=4cm]{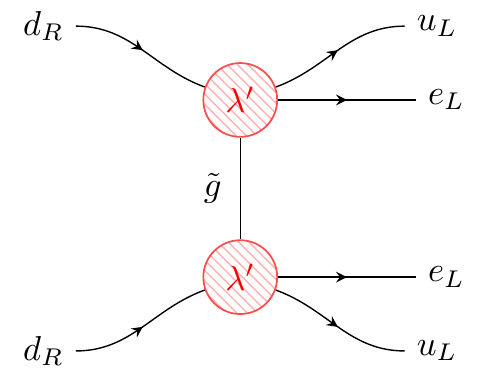}
	\caption{Short-range $0\nu\beta\beta$ decay mechanism induced by the RPV coupling $\lambda'_{111}$. The process can proceed via the exchange of a neutralino $\tilde{\chi}^0_{i}$ or gluino $\tilde{g}$. The  effective interactions indicated by blue and red blobs are induced by the exchange of a selectron, up or down squark (shown in Fig.~\ref{fig:RPV_SUSY_2}). If the neutralino is lighter than the relevant momentum exchange, this mechanism becomes long-range.}
	\label{fig:RPV_SUSY}
\end{figure}
\begin{figure}[t]
	\centering
	\includegraphics[width=3.2cm]{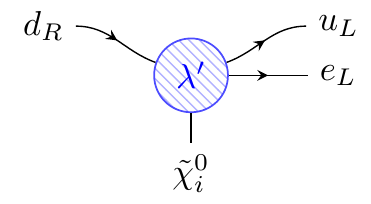} 
	\put(5,25){\normalsize{$ = $}}
	\hspace{1em}
	\includegraphics[width=2.2cm]{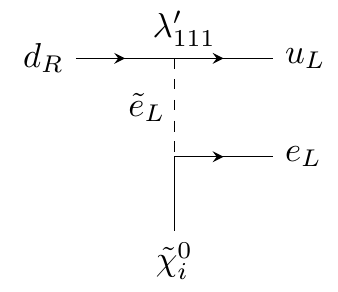} 
	\put(5,25){\normalsize{$ + $}}
	\hspace{1em}
	\includegraphics[width=2.25cm]{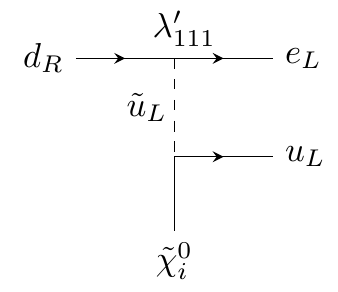} 
	\put(5,25){\normalsize{$ + $}}
	\hspace{1.5em}
	\includegraphics[width=2.6cm]{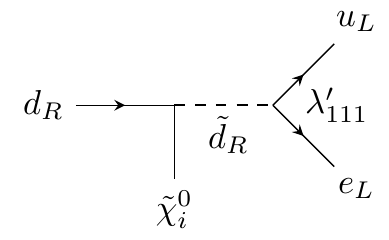} 
	
	\includegraphics[width=3.2cm]{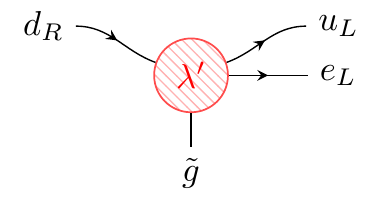} 
	\put(5,25){\normalsize{$ = $}}
	\hspace{1em}
	\includegraphics[width=2.25cm]{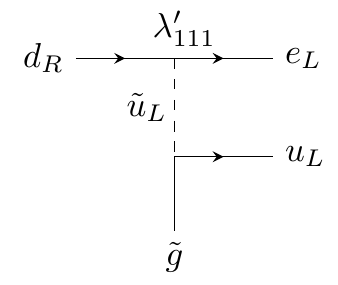} 
	\put(5,25){\normalsize{$ + $}}
	\hspace{1.5em}
	\includegraphics[width=2.6cm]{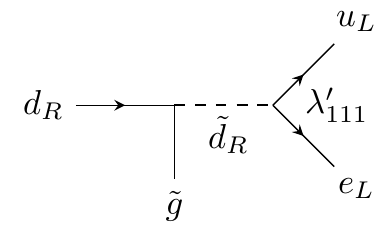}
	\caption{{\it Top:} Diagrams contributing towards the effective interactions at the vertices of the neutralino exchange diagrams in Fig.~\ref{fig:RPV_SUSY}. {\it Bottom:} Similar diagrams contributing towards the gluino exchange effective interactions.}
	\label{fig:RPV_SUSY_2}
\end{figure}

The contribution of the trilinear RPV coupling $\lambda'_{111}$ to $0\nu\beta\beta$ decay is depicted in Fig.~\ref{fig:RPV_SUSY}. The neutralino-mediated diagram can proceed by the exchange of a selectron, up or down squark at the effective interaction vertices, shown in Fig.~\ref{fig:RPV_SUSY_2}, while the gluino-mediated diagram can only take place for the exchange of color-charged up or down squarks. This effective approach is justified because selectrons, up and down squarks with masses less than $p_{\text{F}}\sim 100$ MeV are excluded, as seen in Table~\ref{tab:SUSYlimits}. If the neutralinos and gluinos are also heavier than $p_{\text{F}}\sim 100$ MeV, the process can be described by the dimension-9 effective Lagrangian (using the notation of Ref.~\cite{Pas:2000vn})
\begin{align}
\label{eq:0vbb_shortrange}
\begin{aligned}
\mathcal{L}_{9}&=\frac{G_{\text{F}}^{2}\cos^2\theta_{\text{C}}}{2m_{p}} \left(\epsilon^{RRL}_{1}J_{R} J_{R} + \epsilon^{RRL}_2 J_{R}^{\mu \nu} J_{R \mu \nu}\right)j_{L} + \mathrm{h.c.}\,,
\end{aligned}
\end{align}
where $j_L$, $J_R$ and $J^{\mu\nu}_R$ are the scalar leptonic, scalar and tensor quark currents
\begin{align}
\begin{aligned}
j_{L}=\bar{e}\left(1+\gamma_{5}\right) e^{c}\,,\quad J_{R} =\bar{u}\left(1+\gamma_{5}\right) d\,, \quad J_{R}^{\mu \nu} =\bar{u} \sigma^{\mu \nu}\left(1+\gamma_{5}\right) d\,,
\end{aligned}
\end{align}
respectively, with $\sigma_{\mu\nu} = \frac{i}{2}[\gamma_{\mu},\gamma_{\nu}]$, $G_{\text{F}}$ being the Fermi constant,  $\theta_{\text{C}}$ the Cabibbo mixing angle and $m_p$ the proton mass. To get the Lagrangian in this form one must first perform the appropriate Fierz rearrangements of the squark exchange diagrams in Fig.~\ref{fig:RPV_SUSY_2}. The quark currents must also be color singlets, extracted from the product of two color-triplet and color-antitriplet quark fields~\cite{Allanach:2009xx}. After these necessary steps, the scalar and tensor coefficients $\epsilon^{RRL}_{1}$ and $\epsilon^{RRL}_2$, normalised to the Fermi constant squared $G_{\text{F}}^2$, can be decomposed as
\begin{align}
\begin{aligned}
\epsilon^{RRL}_{1} &= \eta_{\tilde{g}}+\eta_{\tilde{g}}^{\prime}+\eta_{\tilde{\chi}}+\eta_{\tilde{\chi} \tilde{e}}+\eta_{\tilde{\chi} \tilde{f}}\,,\quad\epsilon^{RRL}_{2}&=-\frac{1}{4}\left(\eta_{\tilde{g}}+\eta_{\tilde{\chi}}\right)\,,
\end{aligned}
\end{align}
where the terms,
\begin{align}
\label{eq:etas}
\begin{aligned}
\eta_{\tilde{g}} &=\frac{\pi \alpha_{s}}{6} \frac{\lambda_{111}^{\prime 2}}{G_{\beta}^{2}} \frac{m_{p}}{m_{\tilde{g}}}\left[\frac{1}{m_{\tilde{u}_{L}}^{4}}+\frac{1}{m_{\tilde{d}_{R}}^{4}}-\frac{1}{2 m_{\tilde{u}_{L}}^{2} m_{\tilde{d}_{R}}^{2}}\right]\,, \\
\eta_{\tilde{g}}^{\prime} &=\frac{2 \pi \alpha_{s}}{3} \frac{\lambda_{111}^{\prime 2}}{G_{\beta}^{2}} \frac{m_{p}}{m_{\tilde{g}}} \frac{1}{m_{\tilde{u}_{L}}^{2} m_{\tilde{d}_{R}}^{2}}\,,
\end{aligned}
\end{align}
describe the gluino-mediated diagrams in Fig.~\ref{fig:RPV_SUSY} and
\begin{align}
\label{eq:etas2}
\begin{aligned}
\eta_{\tilde{\chi}} &=\frac{\pi \alpha_{2}}{2} \frac{\lambda_{111}^{\prime 2}}{G_{\beta}^{2}} \sum_{i=1}^{4} \frac{m_{p}}{m_{\tilde{\chi}^0_{i}}}\Bigg[\frac{\big(V_{u\tilde{\chi}^0_{i} }^{LL}\big)^2}{m_{\tilde{u}_{L}}^{4}}+\frac{\big(V_{d\tilde{\chi}^0_{i}}^{RR*}\big)^2}{m_{\tilde{d}_{R}}^{4}}-\frac{V^{LL}_{u\tilde{\chi}^0_{i}} V^{RR*}_{d\tilde{\chi}^0_{i}}}{m_{\tilde{u}_{L}}^{2} m_{\tilde{d}_{R}}^{2}}\Bigg]\,, \\
\eta_{\tilde{\chi} \tilde{f}} &=\pi \alpha_{2} \frac{\lambda_{111}^{\prime 2}}{G_{\beta}^{2}} \sum_{i=1}^{4} \frac{m_{p}}{m_{\tilde{\chi}^0_{i}}}\Bigg[\frac{V^{LL}_{u\tilde{\chi}^0_{i}} V^{RR*}_{d\tilde{\chi}^0_{i}}}{m_{\tilde{u}_{L}}^{2} m_{\tilde{d}_{R}}^{2}}-\frac{V^{LL}_{e\tilde{\chi}^0_{i}}V^{LL}_{u\tilde{\chi}^0_{i}}}{m_{\tilde{e}_{L}}^{2}m_{\tilde{u}_{L}}^{2} }-\frac{V^{LL}_{e\tilde{\chi}^0_{i}} V^{RR*}_{d\tilde{\chi}^0_{i}}}{m_{\tilde{e}_{L}}^{2} m_{\tilde{d}_{R}}^{2}}\Bigg]\,, \\
\eta_{\tilde{\chi} \tilde{e}} &=2 \pi \alpha_{2} \frac{\lambda_{111}^{\prime 2}}{G_{\beta}^{2}} \sum_{i=1}^{4} \frac{m_{p}}{m_{\tilde{\chi}^0_{i}}}\frac{\big(V^{LL}_{e\tilde{\chi}^0_i}\big)^2}{m_{\tilde{e}_{L}}^{4}}\,, \\
\end{aligned}
\end{align}
the neutralino-mediated diagrams. Here, $G_{\beta} = G_{\text{F}}\cos\theta_{\text{C}}$,  $\alpha_2 = g^2/4\pi$ and $\alpha_s = g_s^2/4\pi$, where $g_s$ is the $\mathrm{SU}(3)_c$ coupling constant. For the factors in Eq.~\eqref{eq:etas2} we have neglected the mixing between left- and right-handed first generation sfermions and thus only the factors $V^{LL}_{e\tilde{\chi}^0_{i}}$, $V^{LL}_{u\tilde{\chi}^0_{i}}$ and $V^{RR}_{d\tilde{\chi}^0_{i}}$ appear.

\begin{table}[t]
	\centering
	\renewcommand{\arraystretch}{1.25}
	\setlength\tabcolsep{1.4pt}
	\begin{tabular}{c|ccc|cc|ccc|c|c}
		\hline
		\multirow{3}{*}{Isotope} & \multicolumn{5}{c|}{NMEs}  & & PSFs~\cite{Kotila:2012zza} & &\multicolumn{2}{c}{$T^{\text{exp}}_{1/2}$ $[\mathrm{yr}]$} \\ \cline{2-7} \cline{8-9} \cline{10-11}
		&\multicolumn{3}{c|}{Light~\cite{Kotila:2021xgw}} & \multicolumn{2}{c|}{Heavy~\cite{Deppisch:2020ztt}} & &\multirow{2}{*}{$G^{(0)}_{11+}$ $[\mathrm{yr}^{-1}]$}&& \multirow{2}{*}{Current} & \multirow{2}{*}{Future} \\ \cline{2-4} \cline{5-6}
		&$\mathcal{M}_{\nu}$ & $\mathcal{M}^{XX}_{1}$ & $\mathcal{M}^{XX}_{2}$ & $\mathcal{M}^{XX}_{1}$ & $\mathcal{M}^{XX}_{2}$ &&&&&\\ \hline
		$^{76}$Ge & $-6.64$ & $133$ & $-11.8$ &  $5300$ & $-347$ & & $2.360 \times 10^{-15}$ & & $1.8\times 10^{26}$~\cite{Agostini:2020xta} & $1.3\times 10^{28}$~\cite{LEGEND:2021bnm}  \\
		$^{82}$Se & $-5.46$ & $104$ & $-9.56$ &  $4030$ & $-287$ & & $1.019\times 10^{-14} $ & &$3.5\times 10^{24}$~\cite{Azzolini:2019tta} & $1.0\times 10^{26}$~\cite{Waters:2017wzp}\\
		$^{100}$Mo & $-5.27$ & $214$ & $-10.7$ &  $12400$ & $-377$ & & $1.591\times 10^{-14}$ & &$1.5\times 10^{24}$~\cite{Armengaud:2020luj}& $1.0\times 10^{27}$~\cite{CUPIDInterestGroup:2019inu} \\
		$^{136}$Xe & $-3.60$ & $74.6$ & $-6.21$ &  $3210$ & $-192$ & & $1.456\times 10^{-14}$ & & $1.1\times 10^{26}$~\cite{KamLAND-Zen:2016pfg}& $1.4\times 10^{28}$~\cite{nEXO:2021ujk} \\ \hline
	\end{tabular}
	\caption{Values of the light and heavy NMEs calculated in the IBM-2 model for four different $0\nu\beta\beta$ decay isotopes. For neutralino exchange we require light and heavy scalar and tensor NMEs $\mathcal{M}^{XX}_{1}$ and $\mathcal{M}^{XX}_{2}$ (assuming equal quark current chiralities, $XX=RR$) respectively, while for active neutrino exchange we require the light NMEs $\mathcal{M}_{\nu}$. We also show the relevant phase space factor $G^{(0)}_{11+}$  (identical for short and long-range mechanisms) and the current and future lower bounds on the $0\nu\beta\beta$ decay half-life $T^{\text{exp}}$ for each isotope.}
	\label{tab:SUSYNMEs}
\end{table}

In order to compute the total rate and half-life of the short-range $0\nu\beta\beta$ decay process, the quark currents in Eq.~\eqref{eq:0vbb_shortrange} must be matched onto non-relativistic nucleon currents, detailed in Ref.~\cite{Graf:2018ozy}. To go the nuclear level, a number of approximations simplify the calculation. These are to assume a $0^{+}\rightarrow 0^{+}$ transition, the closure and impulse approximations and $s$-wave final state electrons. The resulting \textit{heavy} nuclear matrix elements (NMEs) have been computed for the scalar and tensor current combinations in Eq.~\eqref{eq:0vbb_shortrange} in the interacting boson model (IBM-2)~\cite{Deppisch:2020ztt}. We list the values of these heavy NMEs, $\mathcal{M}_{1}^{XX}$ (scalar) and $\mathcal{M}_{2}^{XX}$ (tensor), for four different isotopes in Table~\ref{tab:SUSYNMEs}. One must also integrate over the energy and angle of one of the outgoing $s$-wave electrons. The final expression for the $0\nu\beta\beta$ decay half-life is then
\begin{align}
\label{eq:ThalfSUSY}
(T^{0\nu}_{1/2})^{-1} = G^{(0)}_{11+}\left|\epsilon_{\nu}\mathcal{M}_\nu+\epsilon^{RRL}_{1}\mathcal{M}^{RR}_{1}+\epsilon^{RRL}_{2}\mathcal{M}^{RR}_{2}\right|^2\,,
\end{align}
where $G^{(0)}_{11+}$ is the Phase-Space Integral (PSF)
\begin{align}
G_{11+}^{(0)}=\frac{G_{\beta}^4 m_e^2}{32\pi^5R_{A}^{2}\ln 2} \int_{m_{e}}^{Q_{\beta \beta}+m_{e}} d E_{1} E_1E_2p_1p_2 f_{11+}^{(0)}\,,
\end{align}
and $f_{11+}^{(0)}$ is the following combination of radial electron wave functions evaluated at the nuclear surface $r=R_A$,
\begin{align}
f_{11+}^{(0)} = &|g_{-1}(E_1)g_{-1}(E_2)|^2+|g_{-1}(E_1)f_{1}(E_2)|^2 \nonumber\\
&+|f_{1}(E_1)g_{-1}(E_2)|^2+|f_{1}(E_1)f_{1}(E_2)|^2\,.
\end{align}
The values of the PSFs for the four isotopes in Table~\ref{tab:SUSYNMEs} were computed in Ref.~\cite{Kotila:2012zza} using numerical methods. 

In Eq.~\eqref{eq:ThalfSUSY} we also account for the exchange of light neutrinos by including the term $\epsilon_{\nu}\mathcal{M}_{\nu}$, where $\epsilon_{\nu}\equiv m_{\beta\beta}/m_e$ and $\mathcal{M}_{\nu}$ is the standard \textit{light} neutrino exchange NME. The mass $m_{\beta\beta}$ is the standard effective neutrino mass probed in $0\nu\beta\beta$ decay. For simplicity, we consider only the exchange of a single neutrino, such that $m_{\beta\beta}=m_1$. This mass gets a contribution at tree-level $m^{\mathrm{tree}}_1$, but the presence of $\lambda'_{111}$ also induces a contribution to $m_1$ at one-loop~\cite{Hall:1983id, Babu:1989px}. We therefore write
\begin{align}
\label{eq:mbetameta}
m_{\beta\beta} \approx m^{\text{tree}}_1 e^{i\phi_1} + \frac{3}{8 \pi^{2}}\lambda_{111}^{\prime 2} m^2_{d} \frac{A^{d}-\mu \tan \beta}{m_{\tilde{d}_R}^{2}}\,,
\end{align}
where we have assumed degenerate mass for the left- and right-handed down squarks. A general Majorana phase multiplies the tree-level mass, while the loop contribution has the same dependence as $\epsilon_1^{RRL}$ and $\epsilon_2^{RRL}$ on $\lambda_{111}'$ (and so any phase contained in $\lambda_{111}'$ is common between these terms). The $\mathcal{M}^{RR}_1$ and $\mathcal{M}^{RR}_2$ NMEs are positive and negative respectively for the isotopes of interest in Table~\ref{tab:SUSYNMEs}. However, there is a negative sign within $\epsilon_{2}^{RRL}$ and so these contributions add constructively in Eq.~\eqref{eq:ThalfSUSY}. The $\mathcal{M}_\nu$ NMEs are also negative; for $\phi_1 = 0$ ($\pi$), the tree-level light neutrino exchange contribution therefore adds destructively (constructively) with the $\epsilon_1^{RRL}$ and $\epsilon_2^{RRL}$ contributions. We find that the one-loop contribution in Eq.~\eqref{eq:mbetameta} is suppressed with respect to the $\epsilon_1^{RRL}$ and $\epsilon_2^{RRL}$ coefficients by the ratio of down quark to right-handed down squark masses squared (for $m_{\tilde{d}_R}\sim 2$~TeV). Thus, the main constructive or destructive interference arises from the tree-level mass, which can be at most $m_{1}^{\text{tree}}\sim 0.1$~eV from the cosmological limit on the sum of neutrino masses $\sum m_\nu$~\cite{Planck:2018vyg}. For simplicity, we omit the light neutrino exchange mechanism when deriving limits on $\lambda_{111}'$ in the next section. Possible interference effects have been studied in Refs.~\cite{Faessler:2011qw,Meroni:2012qf}. Here, we also neglect the possible contribution of the bilinear RPV interaction to $0\nu\beta\beta$ decay, which has been studied for heavy neutralinos in Refs.~\cite{Hirsch:1998kc,Hirsch:2000jt}.

As shown in Refs.~\cite{Mahajan:2013ixa,Gonzalez:2015ady,Arbelaez:2016zlt,Arbelaez:2016uto,Cirigliano:2017djv,Cirigliano:2018yza,Gonzalez:2017mcg,Liao:2019gex,Ayala:2020gtv}, it is also important to consider the leading-order QCD corrections to Eq.~\eqref{eq:0vbb_shortrange}. These are one-loop diagrams where gluons connect the incoming and outgoing quarks in Fig.~\ref{fig:RPV_SUSY}. These induce an RGE running of the coefficients $\epsilon^{RRL}_{1}$ and $\epsilon^{RRL}_{2}$ from the scale of new physics, $\Lambda_{\mathrm{NP}}$, to the lower limit of perturbative QCD, $\Lambda_{\mathrm{QCD}}\sim 1$~GeV. For example, running from the scale $\Lambda_{\mathrm{NP}}\sim 1$ TeV modifies Eq.~\eqref{eq:ThalfSUSY} to the QCD-improved result
\begin{align}
\label{eq:ThalfSUSY2}
\hspace{-0.4em}(T^{0\nu}_{1/2})^{-1} = G^{(0)}_{11+}\left|\epsilon_{\nu}\mathcal{M}_\nu+\epsilon^{RRL}_{1}\beta^{RR}_{1}+\epsilon^{RRL}_{2}\beta^{RR}_{2}\right|^2\,,
\end{align}
where a mixing between the NMEs now takes place:
\begin{align}
    \begin{pmatrix}
	\beta^{RR}_1 \\ 
	\beta^{RR}_2
	\end{pmatrix}= \begin{pmatrix}
	2.39 & ~-3.83 \\ 
	0.02 & ~0.35
	\end{pmatrix}\begin{pmatrix}
	\mathcal{M}^{RR}_1 \\ 
	\mathcal{M}^{RR}_2 \end{pmatrix}\, .
\end{align}
However, the $0\nu\beta\beta$ decay process takes place at the Fermi scale, $p_{\text{F}}\sim 100$ MeV, where the QCD running of short-range operators is no longer reliable. Ref.~\cite{Gonzalez:2017mcg} investigated naively extrapolating the perturbative results to the sub-GeV regime by assuming the \textquoteleft freezing' of the QCD coupling constant $\alpha_s$. The extrapolation to low energies was not found to modify the perturbative results appreciably, therefore Eq.~\eqref{eq:ThalfSUSY2} will be used in what follows.

In Eqs.~\eqref{eq:etas} and \eqref{eq:etas2} we have only considered the lightest neutralino with mass $m_{\tilde{\chi}^0_1}\gg p_{\text{F}}$. As explained in Section~\ref{sec:MSSMconventions} however, it is always possible to choose (experimentally allowed) values of the MSSM parameters such that the lightest neutralino is very light or massless. In the case $m_{\tilde{\chi}^0_1}\ll p_{\text{F}}$, the exchange of the lightest neutralino contributes towards a long-range $0\nu\beta\beta$ decay mechanism. To take this into account, we make the following replacement in Eq.~\eqref{eq:etas2},
\begin{align}
\label{eq:SUSYinterp}
\frac{1}{m_{\tilde{\chi}^0_1}}\rightarrow\frac{m_{\tilde{\chi}^0_1}}{\braket{\mathbf{p}^2}_{1,2}+m^2_{\tilde{\chi}^0_1}} \approx 
\begin{cases}
\frac{m_{\tilde{\chi}^0_1}}{\braket{\mathbf{p}^2}_{1,2}} & m^2_{\tilde{\chi}^0_1}\ll \braket{\mathbf{p}^2}_{1,2}\\
\frac{1}{m_{\tilde{\chi}^0_1}} & m^2_{\tilde{\chi}^0_1}\gg \braket{\mathbf{p}^2}_{1,2}
\end{cases}
\,,
\end{align}
where the average momentum squared $\braket{\mathbf{p}^2}_1$ is used for the terms $\eta_{\tilde{\chi}}$, $\eta_{\tilde{\chi}\tilde{e}}$ and $\eta_{\tilde{\chi}\tilde{f}}$ in the coefficient $\epsilon_{1}^{RRL}$ (associated with the scalar NME $\mathcal{M}^{RR}_{1}$) and $\braket{\mathbf{p}^2}_2$ is used for the term $\eta_{\tilde{\chi}}$ in the coefficient $\epsilon_{2}^{RRL}$ (associated with the tensor NME $\mathcal{M}^{RR}_{2}$). This function interpolates between the long-range behaviour $m_{\tilde{\chi}^0_1}/\braket{\mathbf{p}^2}_{1,2}$ for $m^2_{\tilde{\chi}^0_1}\ll \braket{\mathbf{p}^2}_{1,2}$ and short-range behaviour $1/m_{\tilde{\chi}^0_1}$ for $m^2_{\tilde{\chi}^0_1}\gg \braket{\mathbf{p}^2}_{1,2}$. The average momentum squared of the exchanged neutralino for scalar and tensor interactions, $\braket{\mathbf{p}^2}_1$ and $\braket{\mathbf{p}^2}_2$, respectively, are given by the following ratio of heavy and light NMEs,
\begin{align}
\label{eq:mominterp}
\braket{\mathbf{p}^2}_{1,2} = m_p m_e \bigg|\frac{(\mathcal{M}_{1,2}^{RR})_{h}}{(\mathcal{M}_{1,2}^{RR})_{l}}\bigg|\,,
\end{align}
where the $h$ and $l$ subscripts indicate the heavy and light NMEs, respectively. Here, the light NMEs for scalar and tensor interactions are taken from Ref. \cite{Kotila:2021xgw}, calculated in the IBM-2 formalism with a long-range neutrino potential (but also applicable for an exchanged neutralino). The values of the light NMEs $\mathcal{M}_{1}^{XX}$ and $\mathcal{M}_{2}^{XX}$ (assuming equal quark current chiralities, $XX=RR$) are given in Table~\ref{tab:SUSYNMEs} for the four isotopes of interest in this work. In Table~\ref{tab:SUSYNMEs2} we subsequently compute the values of $\sqrt{\braket{\mathbf{p}^2}_1}$ and $\sqrt{\braket{\mathbf{p}^2}_2}$ for these isotopes. It can be seen that, as expected, $\sqrt{\braket{\mathbf{p}^2}_{1,2}} \sim p_{\text F} \sim 100$~MeV. However, the average momentum transfers associated with the scalar interaction are slightly larger than those for the tensor interaction, and variation can be seen between the different isotopes.

\begin{table}[t]
	\centering
	\renewcommand{\arraystretch}{1.25}
	\setlength\tabcolsep{2.4pt}
	\begin{tabular}{c|c|c}
		\hline
		Isotope & $\sqrt{\braket{\mathbf{p}^2}_1}$ [MeV] & $\sqrt{\braket{\mathbf{p}^2}_2}$ [MeV] \\\hline
		$^{76}$Ge & $138$ & $119$ \\
		$^{82}$Se & $136$ & $120$ \\
		$^{100}$Mo & $167$ & $130$ \\
		$^{136}$Xe & $144$ & $122$ \\\hline
	\end{tabular}
	\caption{Root mean squares of the neutralino momentum exchange found by inserting the light and heavy scalar and tensor NMEs in Table~\ref{tab:SUSYNMEs} into the formula in Eq.~\eqref{eq:mominterp}. We show the values for the four relevant isotopes used in this work.}
	\label{tab:SUSYNMEs2}
\end{table}
In order to make a more accurate prediction for the change between long- and short-range behaviour, one should \textit{a priori} use the chiral effective field theory formalism of Refs.~\cite{Cirigliano:2017djv, Cirigliano:2018yza, Dekens:2020ttz}, which not only examined the mass dependence of the NMEs as in Eq.~\eqref{eq:SUSYinterp} (due to the propagator of the exchanged particle) but also the chiral low energy constants (LECs). This is particularly important for neutralino masses $m_{\tilde{\chi}^0_1}$ in the range 100~MeV to 1~GeV, where, as mentioned previously, the perturbative QCD corrections to short-range operators are no longer applicable and a more precise calculation is possible beyond the interpolating formula in Eq.~\eqref{eq:SUSYinterp}~\cite{Babic:2018ikc, Dekens:2020ttz}. As detailed in Ref.~\cite{Cirigliano:2018hja}, one must also include leading-order short-range contributions for any long-range $0\nu\beta\beta$ decay mechanism (whether it be active neutrino or light neutralino exchange). However, as this work considers a wider neutralino mass range of 10~keV to 10~TeV, for simplicity we take the replacement Eq.~\eqref{eq:SUSYinterp} in $\epsilon_{1}^{RRL}$ and $\epsilon_{2}^{RRL}$ in Eq.~\eqref{eq:ThalfSUSY2} as a good approximation of the long- and short-range limits. Since we use results from the IBM-2 nuclear structure framework, we neglect any theoretical uncertainties in the NMEs. These are expected to change NMEs by a factor of two or more, $\Delta\mathcal{M}/\mathcal{M}\sim 2-3$, as the dominant source of uncertainty. Because the $0\nu\beta\beta$ decay rate scales as $|\lambda'_{111}|^4$, this results in an uncertainty of $\sqrt{\Delta\mathcal{M}/\mathcal{M}}$ on the limits on $\lambda'_{111}$ we derive below.

\begin{figure}[t!]
	\centering
	\includegraphics[width=0.65\textwidth]{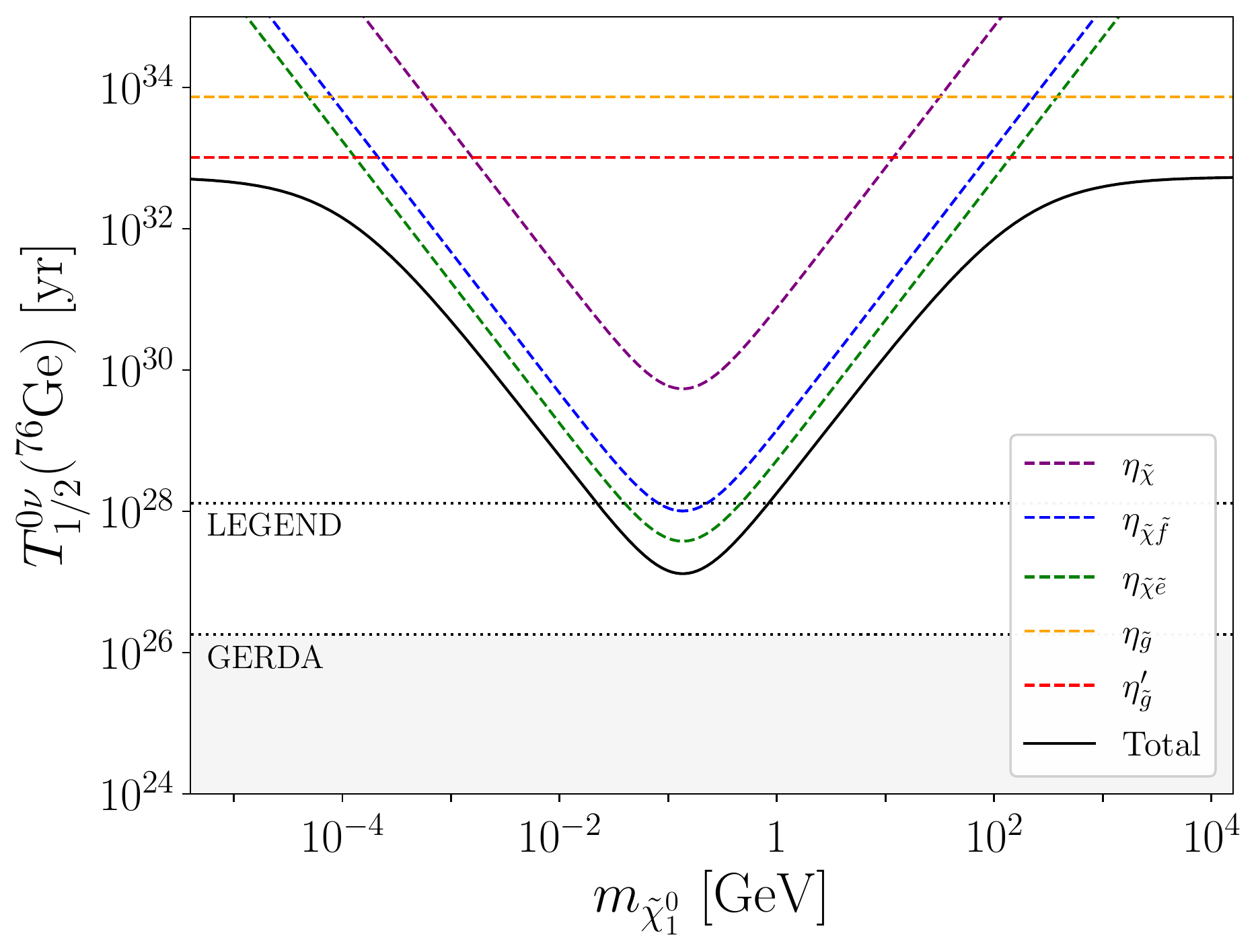}
	\caption{The contributions of the RPV terms $\eta_{\tilde{\chi}}, \eta_{\tilde{\chi}\tilde{f}}, \eta_{\tilde{\chi}\tilde{e}}, \eta_{\tilde{g}}$ and $\eta_{\tilde{g}}'$ to the $0\nu\beta\beta$ decay half-life of $^{76}$Ge as a function of the lightest neutralino mass $m_{\tilde{\chi}^0_1}$. The values $\lambda'_{111}=10^{-3}$, $m_{\tilde{e}_L}=m_{\tilde{u}_L}=m_{\tilde{d}_R}=2$~TeV and $m_{\tilde{g}}=2.5$~TeV are chosen for the other MSSM parameters. We compare the half-life to the current reach of GERDA~\cite{Agostini:2020xta} and future sensitivity of LEGEND~\cite{LEGEND:2021bnm}, shown by the horizontal dotted lines. }
	\label{fig:eta_plot}
\end{figure}

A few points are in order before deriving constraints on the parameter space of the RPV coupling $\lambda'_{111}$ and the relevant MSSM masses $m_{\tilde{\chi}^0_1}$, $m_{\tilde{e}_{L}}$, $m_{\tilde{u}_{L}}$, $m_{\tilde{d}_{R}}$ and $m_{\tilde{g}}$. Previous analyses have noted the so-called \textit{gluino dominance} for a region of the MSSM parameter space~\cite{Hirsch:1995ek}. This is the observation that the gluino exchange terms in Eq.~\eqref{eq:etas} are larger than neutralino exchange terms in Eq.~\eqref{eq:etas2}, i.e.
\begin{align}
\label{eq:etascompare}
\eta_{\tilde{g}},~\eta_{\tilde{g}}' \gg \eta_{\tilde{\chi}},~\eta_{\tilde{\chi}\tilde{e}},~\eta_{\tilde{\chi}\tilde{f}}\,,
\end{align}
if, for example, $m_{\tilde{e}_L}\approx m_{\tilde{u}_L}\approx m_{\tilde{d}_R}$ and $m_{\tilde{\chi}^0_{i}}\gtrsim 0.05 \,m_{\tilde{g}}$. This is shown explicitly in Fig.~\ref{fig:eta_plot}, which plots the contribution of each $\eta$ term (and also the total contribution) to the $0\nu\beta\beta$ decay half-life as a function of $m_{\tilde{\chi}^0_1}$, for $m_{\tilde{e}_L} =  m_{\tilde{u}_L} = m_{\tilde{d}_R} = 2$~TeV and $m_{\tilde{g}}=2.5$~TeV, and taking $^{76}$Ge isotope for illustration. We can see that the neutralino contributions $\eta_{\tilde{\chi}},~\eta_{\tilde{\chi}\tilde{e}},~\eta_{\tilde{\chi}\tilde{f}}$ to the $0\nu\beta\beta$ decay rate indeed become smaller than the gluino contributions $\eta_{\tilde{g}},~\eta_{\tilde{g}}'$ above $m_{\tilde{\chi}^0_1}\sim 0.05 \, m_{\tilde{g}}$, defining a region of gluino dominance. Below $m_{\tilde{\chi}^0_1}\sim 0.05 \, m_{\tilde{g}}$ is instead a region of \textit{neutralino dominance}, which was so far overlooked in the literature. Due to the interpolating formula in Eq.~\eqref{eq:SUSYinterp}, the neutralino contributions reach a maximum at $m_{\tilde{\chi}^0_1}\sim \sqrt{\braket{\mathbf{p}^2}}_{1,2}$ and then start to decrease, eventually falling below the gluino contributions and giving gluino dominance for $m_{\tilde{\chi}^0_1}\lesssim 5\times10^{-8}\,m_{\tilde{g}}$.

However, it should be stated that the dominating $\eta$ term heavily depends on the choice of  superpartner masses. In Fig.~\ref{fig:eta_plot}, it can be seen that for $m_{\tilde{e}_L} =  m_{\tilde{u}_L} = m_{\tilde{d}_R} = 2$~TeV and $m_{\tilde{g}}=2.5$~TeV, the contribution from $\eta_{\tilde{\chi}\tilde{e}}$ is always larger than $\eta_{\tilde{\chi}}$ and $\eta_{\tilde{\chi}\tilde{f}}$. Furthermore, for large $m_{\tilde{\chi}^0_1}$ the collider constraints on the selectron mass are relaxed. For $m_{\tilde{e}_L} = 90$~GeV, $m_{\tilde{u}_L} = m_{\tilde{d}_R} = 2$~TeV and $m_{\tilde{g}} = 2.5$~TeV, the contribution from $\eta_{\tilde{\chi}\tilde{e}}$ is increased considerably, dominating over $\eta_{\tilde{g}}$ and $\eta_{\tilde{g}}'$ up to $m_{\tilde{\chi}^0_1}\sim 10^{5}$ TeV. This therefore demonstrates that, while useful in simplified analyses, the assumption of \textit{gluino dominance} is by no means guaranteed.

\section{Excluded Regions in RPV Parameter Space}
\label{sec:RPVexcl}

We will now examine the constraints on the RPV MSSM parameter space relevant to $0\nu\beta\beta$ decay: ($\lambda_{111}',m_{\tilde{\chi}^0_1}, m_{\tilde{e}_{L}}, m_{\tilde{u}_{L}}, m_{\tilde{d}_{R}}, m_{\tilde{g}}$). For simplicity, we will consider only the lightest neutralino with a mass above or below the Fermi scale $p_{\text{F}}$. All other neutralinos are assumed to be much heavier and their contribution neglected.

If an experiment sets a lower bound $T^{\text{exp}}_{1/2}$ on the $0\nu\beta\beta$ decay half-life (as shown in Table.~\ref{tab:SUSYNMEs}), this can be related to the expression in Eq.~\eqref{eq:ThalfSUSY2} with the interpolating formula of Eq.~\eqref{eq:SUSYinterp} for the lightest neutralino. The resulting inequality can then be rearranged for the RPV coupling $\lambda_{111}'$, i.e.
\begin{align}
\label{eq:Thalfcompare}
T^{\text{exp}}_{1/2} < T^{0\nu}_{1/2}\quad\Rightarrow \quad \lambda'_{111} < F(m_{\tilde{\chi}^0_1},m_{\tilde{e}_L},m_{\tilde{u}_L},m_{\tilde{d}_R},m_{\tilde{g}},T^{\text{exp}}_{1/2})\,.
\end{align}
In other words, the non-observation of $0\nu\beta\beta$ decay sets an upper limit on $\lambda_{111}'$ as a function of $T^{\text{exp}}_{1/2}$ and the superpartner masses. On the other hand, as discussed in Section~\ref{sec:RPVconstraints}, large swathes of this parameter space are already excluded by collider experiments. 

\begin{figure}
	\centering
	\includegraphics[width=0.65\textwidth]{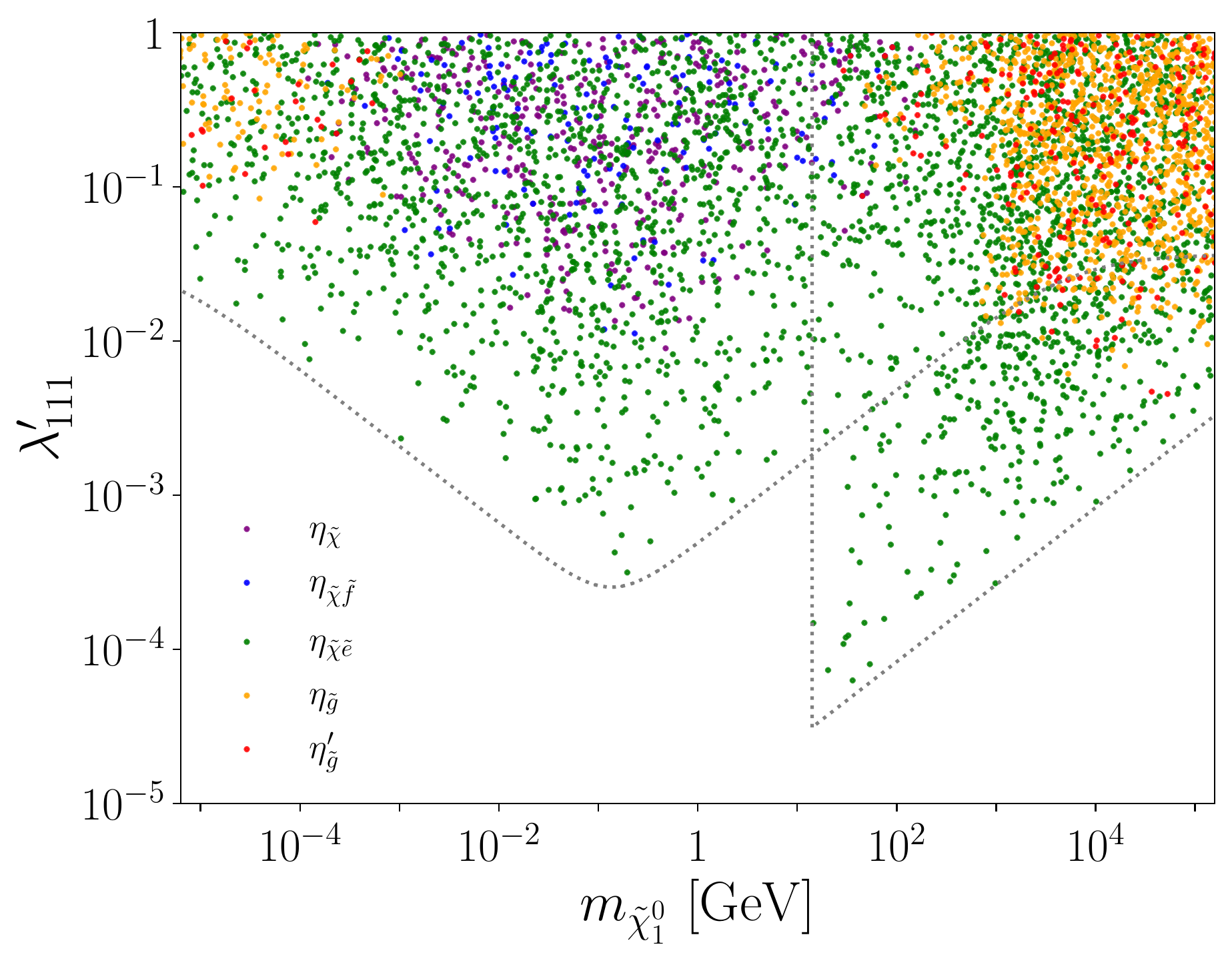}
	\caption{Randomly generated points in the $(\lambda^{\prime}_{111},\, m_{\tilde{\chi}^0_1})$ plane excluded by colliders and the non-observation of $0\nu\beta\beta$ decay in $^{76}$Ge by GERDA, $T^{\text{exp}}_{1/2} \geq T^{0\nu}_{1/2}$. The colours indicate the $\eta$ term that dominates in Eq.~\eqref{eq:etas} or \eqref{eq:etas2}. The two dashed lines indicate the values of $(\lambda^{\prime}_{111},\, m_{\tilde{\chi}^0_1})$ saturating the current experimental lower limit on $T_{1/2}^{0\nu}$.}
	\label{fig:SUSYscan}
\end{figure}

To get an idea of the region of parameter space that is also excluded by $0\nu\beta\beta$ decay, we perform a scan over the region of parameter space that is still currently allowed by collider experiments. To do this, we randomly select sets of parameters in the ranges $\lambda_{111}'\in [10^{-5},\,1]$, $m_{\tilde{e}_L}, m_{\tilde{u}_L}, m_{\tilde{d}_R},m_{\tilde{g}}\in [1,\,10^{5}]$~GeV. A given set is discarded if a combination of any two of the superpartner masses is forbidden by  ATLAS or CMS. We also discard any set containing a superpartner mass below the approximate pMSSM lower bound in Table~\ref{tab:SUSYlimits}. With each randomly selected $\lambda_{111}'$ in an \textquoteleft allowed' set, we now make the comparison
\begin{align}
\label{eq:Thalfscan}
\lambda^{\prime}_{111}\geq F(m_{\tilde{\chi}^0_1},m_{\tilde{e}_L}, m_{\tilde{u}_L}, m_{\tilde{d}_R}, m_{\tilde{g}}, T^{\text{exp}}_{1/2})\,.
\end{align}
If this is condition is met we retain the set as the $0\nu\beta\beta$ decay-excluded point and if not we discard it. With the finally selected sets that are excluded by $0\nu\beta\beta$ decay, we also evaluate the contribution of each of the $\eta$ terms in Eqs.~\eqref{eq:etas} and \eqref{eq:etas2} to the half-life. In Fig.~\ref{fig:SUSYscan}, we plot the excluded combinations of the parameters in the $(\lambda^{\prime}_{111},\, m_{\tilde{\chi}^0_1})$ plane, using the current lower limit on the $^{76}$Ge $0\nu\beta\beta$ decay half-life from GERDA-II~\cite{Agostini:2020xta}. We also colour each point according to the $\eta$ term that dominates, i.e. $\eta_{\tilde{\chi}}$ (purple), $\eta_{\tilde{\chi}\tilde{f}}$ (blue), $\eta_{\tilde{\chi}\tilde{e}}$ (green), $\eta_{\tilde{g}}$ (orange) and $\eta_{\tilde{g}}'$ (red).

\begin{figure}
	\centering
	\includegraphics[width=0.65\textwidth]{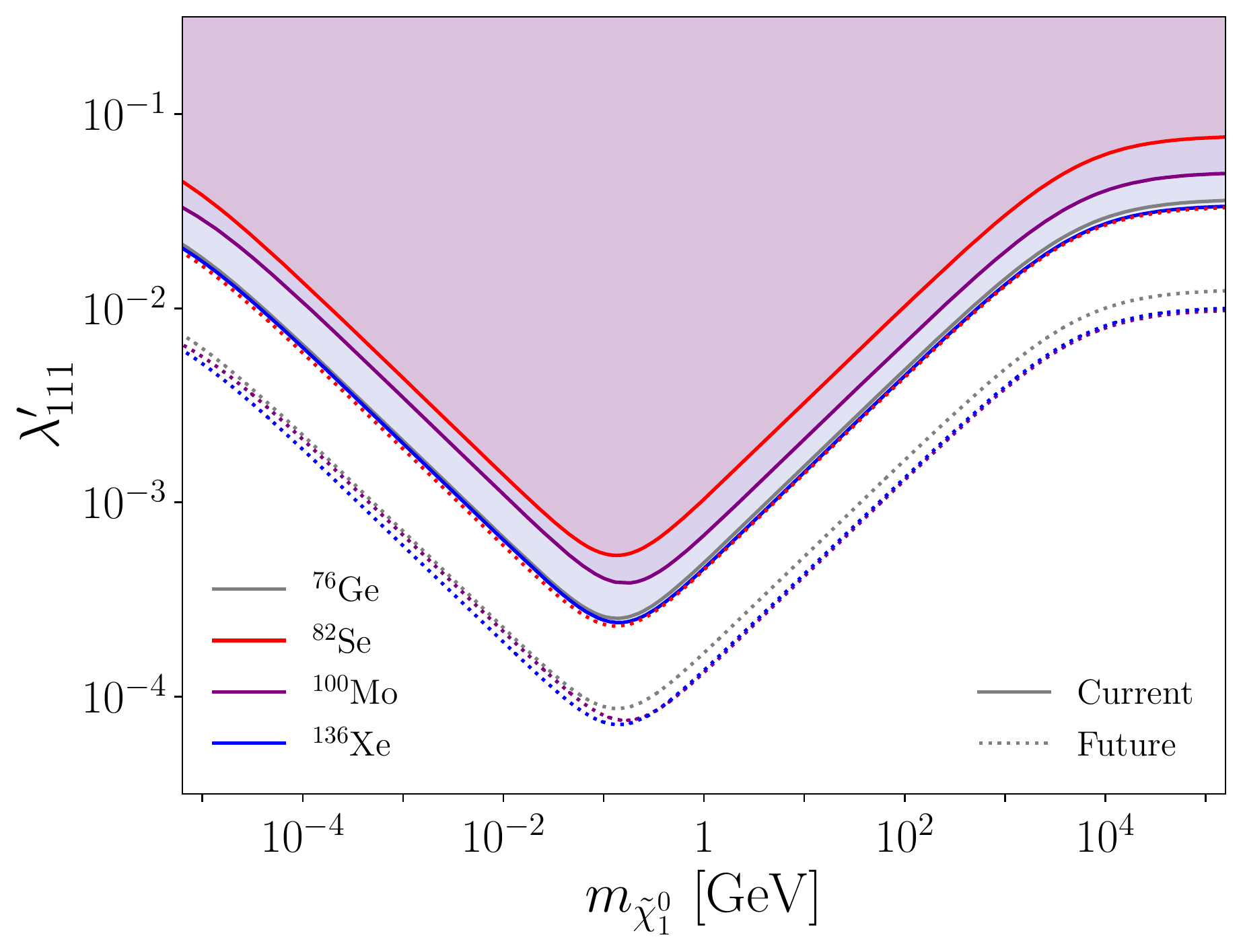}
	\caption{Upper limits on $\lambda_{111}'$ from the current lower limits on the $0\nu\beta\beta$ decay half-life of $^{76}$Ge (GERDA), $^{82}$Se (CUPID), $^{100}$Mo (CUPID) and $^{136}$Xe (KamLAND-Zen) for the benchmark superpartner masses $m_{\tilde{e}_L} = 700$~GeV, $m_{\tilde{u}_L} = m_{\tilde{d}_R} = 1900$~GeV and $m_{\tilde{g}} = 2350$~GeV. The corresponding limits from the future projections are also shown by the dashed lines.}
	\label{fig:SUSYscan2}
\end{figure}

It can be seen that the excluded points extend down to two dashed lines, which show the values of $(\lambda^{\prime}_{111},\, m_{\tilde{\chi}^0_1})$ that saturate the current experimental lower bound on $T_{1/2}^{0\nu}$. Furthermore, the $\eta_{\tilde{\chi}\tilde{e}}$ term (i.e. selectron exchange) dominates for the smallest excluded values of $\lambda'_{111}$ over the whole range of $m_{\tilde{\chi}^0_1}$. The dashed line to the left of the plot corresponds to the case where the selectron, up and down squark and gluino masses take the minimum values allowed by ATLAS and CMS for small neutralino masses. These are $m_{\tilde{e}_L} = 700$~GeV, $m_{\tilde{u}_L} = m_{\tilde{d}_R} = 1900$~GeV and $m_{\tilde{g}} = 2350$~GeV, respectively. The second dashed lines appears at higher values of $m_{\tilde{\chi}^0_1}$, where the selectron can evade the ATLAS and CMS limits and the approximate limit from the pMSSM applies. This dashed line then corresponds to $m_{\tilde{e}_L} = 90$~GeV, $m_{\tilde{u}_L} = m_{\tilde{d}_R} = 1900$~GeV and $m_{\tilde{g}} = 2350$~GeV, but is only valid down to $m_{\tilde{\chi}^0_1}\sim 15$~GeV, where ATLAS begins to exclude $m_{\tilde{e}_L} = 90$~GeV. As the chosen parameters for the first dashed line provide the best limits over a wide range of $m_{\tilde{\chi}^0_1}$, we choose these as benchmark parameters in the following discussion.

\begin{figure}[t!]
	\centering
	\includegraphics[width=0.65\textwidth]{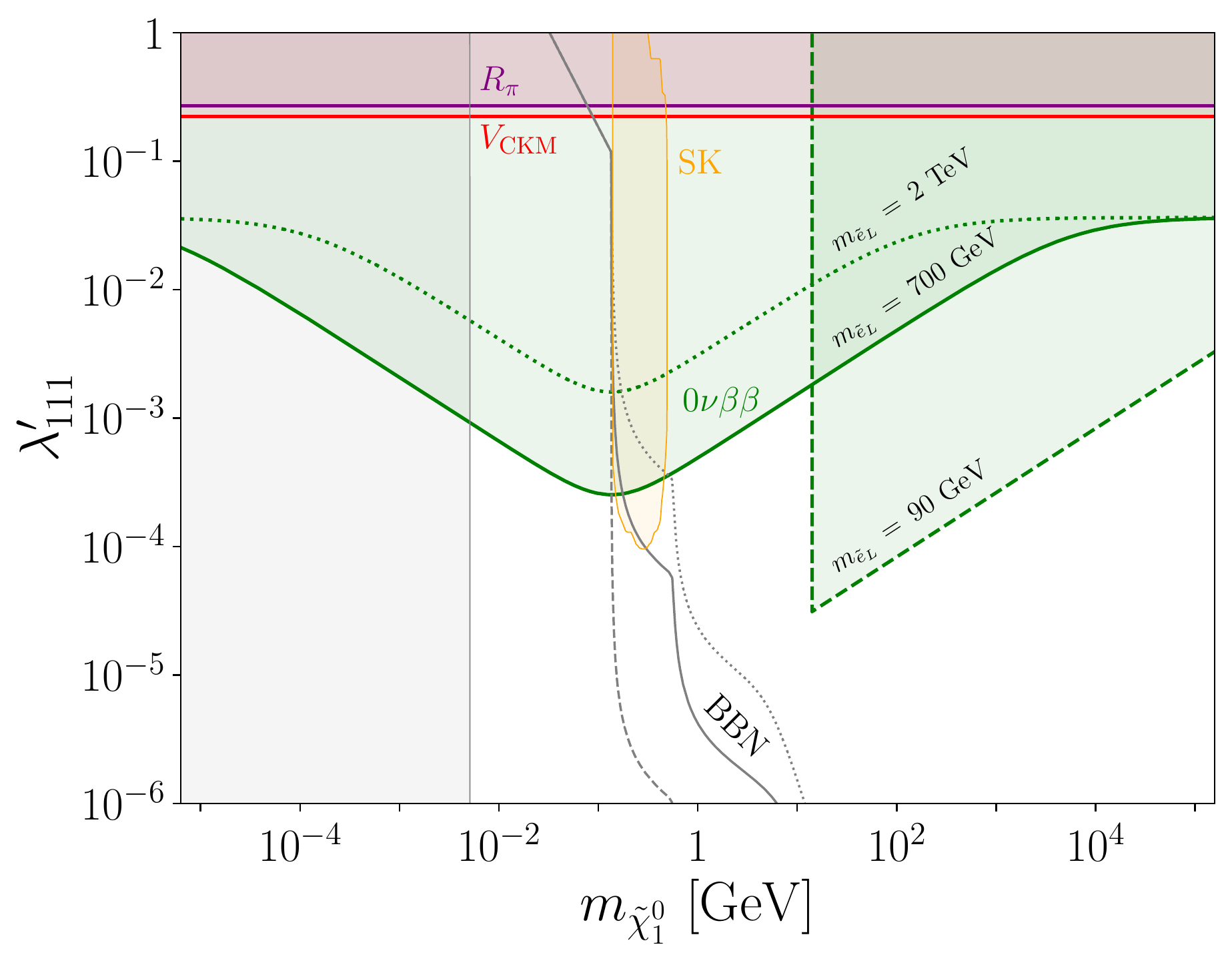}
	\caption{Excluded regions in the $(\lambda^{\prime}_{111},\, m_{\tilde{\chi}^0_1})$ plane. The green regions are excluded from the current lower limit on the $^{76}$Ge $0\nu\beta\beta$ decay half-life for $m_{\tilde{u}_L} = m_{\tilde{d}_R} = 1900$~GeV and $m_{\tilde{g}} = 2350$~GeV and three different values of the selectron mass; $m_{\tilde{e}_L} = 700$~GeV (solid), $m_{\tilde{e}_L} = 90$~GeV (dashed) and $m_{\tilde{e}_L} = 2$~TeV (dotted). The red and purple regions are excluded from measurements of $V_{ud}$ and $R_\pi$ respectively. The orange region is excluded by the non-observation of $\tilde{\chi}^0_1$ decays in Super-Kamiokande, assuming $\lambda'_{112} = \lambda'_{111}$, $m_{\tilde{e}_L} = 700$~GeV and $m_{\tilde{u}_L} = m_{\tilde{d}_R} = 1900$~GeV. The regions to the left of the solid and dashed grey lines are excluded by BBN, i.e. requiring the $\tilde{\chi}^0_1$ lifetime to be less than 1~second. The solid vertical grey line and shaded region applies for a gravitino mass $m_{\tilde{\psi}}$ at the lower bound from L3, whereas the solid, dashed and dotted grey lines correspond to the same three values of $m_{\tilde{e}_L}$ (and benchmark values of $m_{\tilde{u}_L}$, $m_{\tilde{d}_R}$ and $m_{\tilde{g}}$) as the $0\nu\beta\beta$ decay limits, with $m_{\tilde{\psi}}\gg m_{\tilde{e}_L}$.}
	\label{fig:SUSY-constraint-plot}
\end{figure}

In Fig.~\ref{fig:SUSYscan2} we show the excluded regions in the $(\lambda^{\prime}_{111},\, m_{\tilde{\chi}^0_1})$ plane derived from the current and future experimental lower limits on the $0\nu\beta\beta$ decay half-life, using the benchmark choices $m_{\tilde{e}_L} = 700$~GeV, $m_{\tilde{u}_L} = m_{\tilde{d}_R} = 1900$~GeV and $m_{\tilde{g}} = 2350$~GeV. We use the lower limits on the half life for the $^{76}$Ge experiments GERDA~\cite{Agostini:2020xta} (current, solid grey) and LEGEND~\cite{LEGEND:2021bnm} (future, dashed grey), $^{82}$Se experiments CUPID~\cite{Azzolini:2019tta} (current, solid red) and SuperNEMO~\cite{Waters:2017wzp} (future, dashed red), $^{100}$Mo experiment CUPID~\cite{Armengaud:2020luj,CUPIDInterestGroup:2019inu} (current, solid purple; future, dashed purple), and $^{136}$Xe experiments KamLAND-Zen~\cite{KamLAND-Zen:2016pfg} (current, solid blue) and nEXO~\cite{nEXO:2021ujk} (future, dashed blue). Even though the lower bound on $T_{1/2}^{0\nu}$ is expected to be less stringent for $^{100}$Mo compared to $^{76}$Ge in future, the light and heavy scalar NMEs $(\mathcal{M}_1^{RR})_{l}$ and $(\mathcal{M}_2^{RR})_{h}$ for $^{100}$Mo are larger and result in a more stringent future limit on $\lambda_{111}'$. Overall, the $^{136}$Xe limits give the most stringent constraint on $\lambda'_{111}$. It should be noted that for $m_{\tilde{\chi}_1^0}>700$~GeV (to the right of the plot) the selectron becomes the LSP. However, the selectron can still decay via the RPV decay $\tilde{e}_L \rightarrow u\bar{d}$ and is not long-lived.

We will now combine the various limits in the $(\lambda^{\prime}_{111},\, m_{\tilde{\chi}^0_1})$ plane to gauge the ability of $0\nu\beta\beta$ decay experiments to constrain the parameter space. In Fig.~\ref{fig:SUSY-constraint-plot}, we show in green the constraints from the non-observation of $^{76}$Ge $0\nu\beta\beta$ decay in GERDA for different values of the selectron mass; $m_{\tilde{e}_L} = 700$~GeV (solid), $m_{\tilde{e}_L} = 90$~GeV (dashed) and $m_{\tilde{e}_L} = 2$~TeV (dotted), for benchmark values of the other superpartner masses. Similarly, we plot in grey the constraints from the condition that the lightest neutralino lifetime is no longer than 1~second in order to not interfere with BBN; the regions to the left of these lines are excluded. Again, the solid, dashed and dotted lines correspond to the same values of the selectron other superpartner masses as those for $0\nu\beta\beta$ decay. For these lines we have assumed a gravitino mass much larger than the selectron mass $m_{\tilde{\psi}} \gg m_{\tilde{e}_L}$. If we instead set the gravitino mass to be at the lower limit from the L3 experiment, shown in Table~\ref{tab:SUSYlimits}, then the bound shifts to the left and is indicated by the grey shaded region. Finally, in red and purple we show the upper limits on $\lambda_{111}'$ from measurements of $V_{\mathrm{CKM}}$ [cf.~Eq.~\eqref{eq:lamCKM}] and $R_{\pi}$ [cf.~Eq.~\eqref{eq:lamRpi}], respectively, and in orange the upper limit imposed by the non-observation of displaced atmospheric neutralino decays in Super-Kamiokande~\cite{Candia:2021bsl}, assuming $\lambda'_{111} = \lambda'_{112}$ and the benchmark values $m_{\tilde{e}_L} = 700$~GeV, $m_{\tilde{u}_L} = m_{\tilde{d}_R} = 1900$~GeV.

\begin{figure}[t!]
	\centering
	\includegraphics[width=0.48\textwidth]{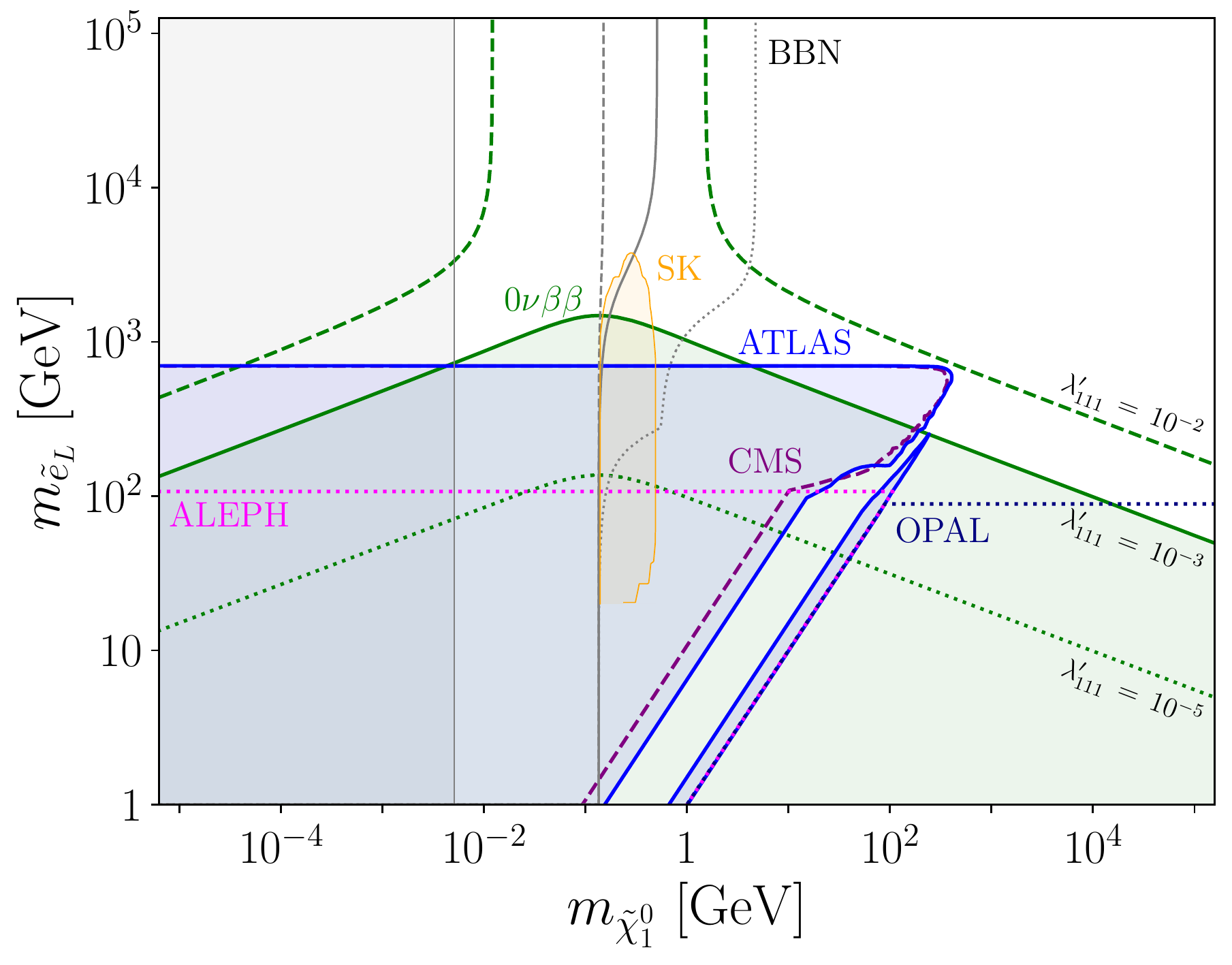}
	\includegraphics[width=0.5\textwidth]{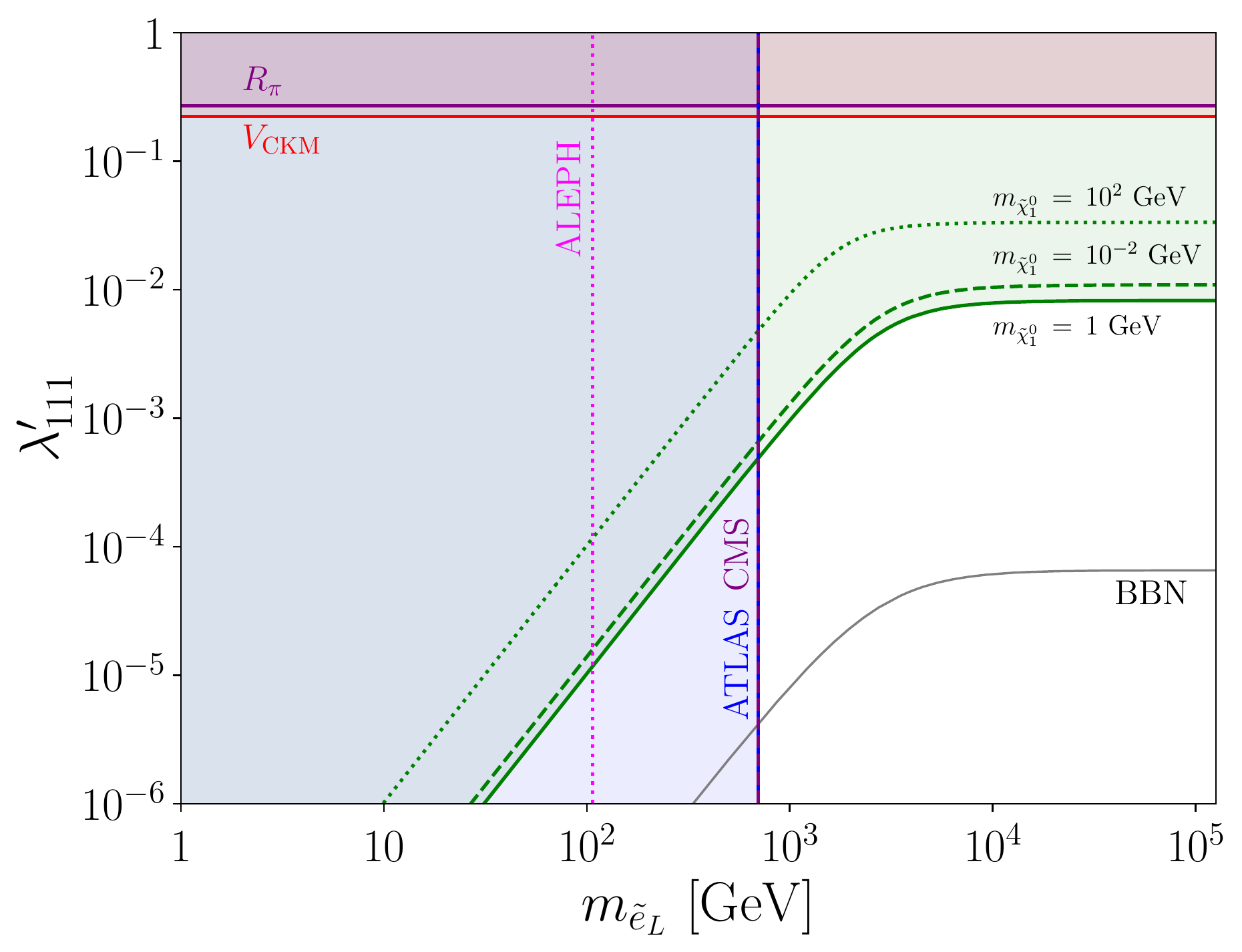}
	\caption{Excluded regions in the $(m_{\tilde{e}_L},\, m_{\tilde{\chi}^0_1})$ plane (left) and $(\lambda^{\prime}_{111},\,m_{\tilde{e}_L})$ plane (right). The green regions are excluded by $0\nu\beta\beta$ decay ($^{76}$Ge) for three different values of $\lambda^{\prime}_{111}$ and $m_{\tilde{\chi}^0_1}$ respectively. The blue, dashed purple, magenta and dark blue regions are excluded by collider constraints. The solid red and purple regions are excluded by measurements of $V_{ud}$ and $R_\pi$ respectively, while the orange region is excluded by the non-observation of $\tilde{\chi}_1^0$ decays in Super-Kamiokande, with $\lambda'_{111} = \lambda'_{112} = 10^{-3}$ and $m_{\tilde{u}_L} = m_{\tilde{d}_R} = 1900$~GeV. The regions to the left and below the grey lines are excluded by BBN, respectively.}
	\label{fig:SUSY-constraint-plot-2}
\end{figure}

It can be seen that the current collider limits of ATLAS and CMS allow $0\nu\beta\beta$ decay to probe a significant portion of the parameter space (smaller values of $\lambda_{111}'$) compared to $V_{\mathrm{CKM}}$ and $R_{\pi}$ measurements. However, the bounds from $0\nu\beta\beta$ decay become weaker as the mass of the selectron (and the other superpartner masses) are increased. The naive limit from BBN (for a large gravitino mass) excludes values of $m_{\tilde{\chi}^0_1}$ below roughly the pion mass, which happens to be where the change from short-range to long-range behaviour in $0\nu\beta\beta$ decay takes place. Furthermore, as one increases the superpartner masses, the naive BBN bounds become \textit{more} stringent. The $0\nu\beta\beta$ decay bound nevertheless depends on one less parameter than the BBN bound, and therefore provide more robust limits for small neutralino masses.

To see the dependence of the $0\nu\beta\beta$ limits on the selectron mass, we plot in the left panel of Fig.~\ref{fig:SUSY-constraint-plot-2} the excluded regions in the $(m_{\tilde{e}_L},\, m_{\tilde{\chi}^0_1})$ plane. The green regions are excluded by $0\nu\beta\beta$ decay ($^{76}$Ge) for three different values of the $\lambda^{\prime}_{111}$. The regions below the blue, purple, magenta and dark blue lines are excluded by ATLAS, CMS, ALEPH and OPAL, respectively~\cite{Aad:2019vnb,Sirunyan:2020eab,Abbiendi:2003rn,Heister:2002jca}. The orange region is again excluded by the non-observation of atmospheric neutralino decays in Super-Kamiokande with $\lambda'_{111} = \lambda'_{112} = 10^{-3}$~\cite{Candia:2021bsl}. The regions to the left of the grey lines are again excluded by BBN, with solid or dashed line depending on the gravitino mass. To the right of Fig.~\ref{fig:SUSY-constraint-plot-2} we show the excluded regions in the $(\lambda^{\prime}_{111},\,m_{\tilde{e}_L})$ parameter space, with green regions again excluded by $0\nu\beta\beta$ decay ($^{76}$Ge) for three different values of $m_{\tilde{\chi}^0_1}$. The red and purple regions are excluded from measurements of $V_{ud}$ and $R_\pi$, respectively, while the blue region is excluded by ATLAS and CMS for $m_{\tilde{\chi}^0_1} = 0\,\,\mathrm{GeV}$. The region to the left of the dashed yellow line is excluded by ALEPH for $m_{\tilde{\chi}^0_1}<m_{\tilde{e}_L}$. The region below the dashed grey line is excluded by BBN for a large gravitino mass. 

\begin{figure}[t!]
	\centering
	\includegraphics[width=0.65\textwidth]{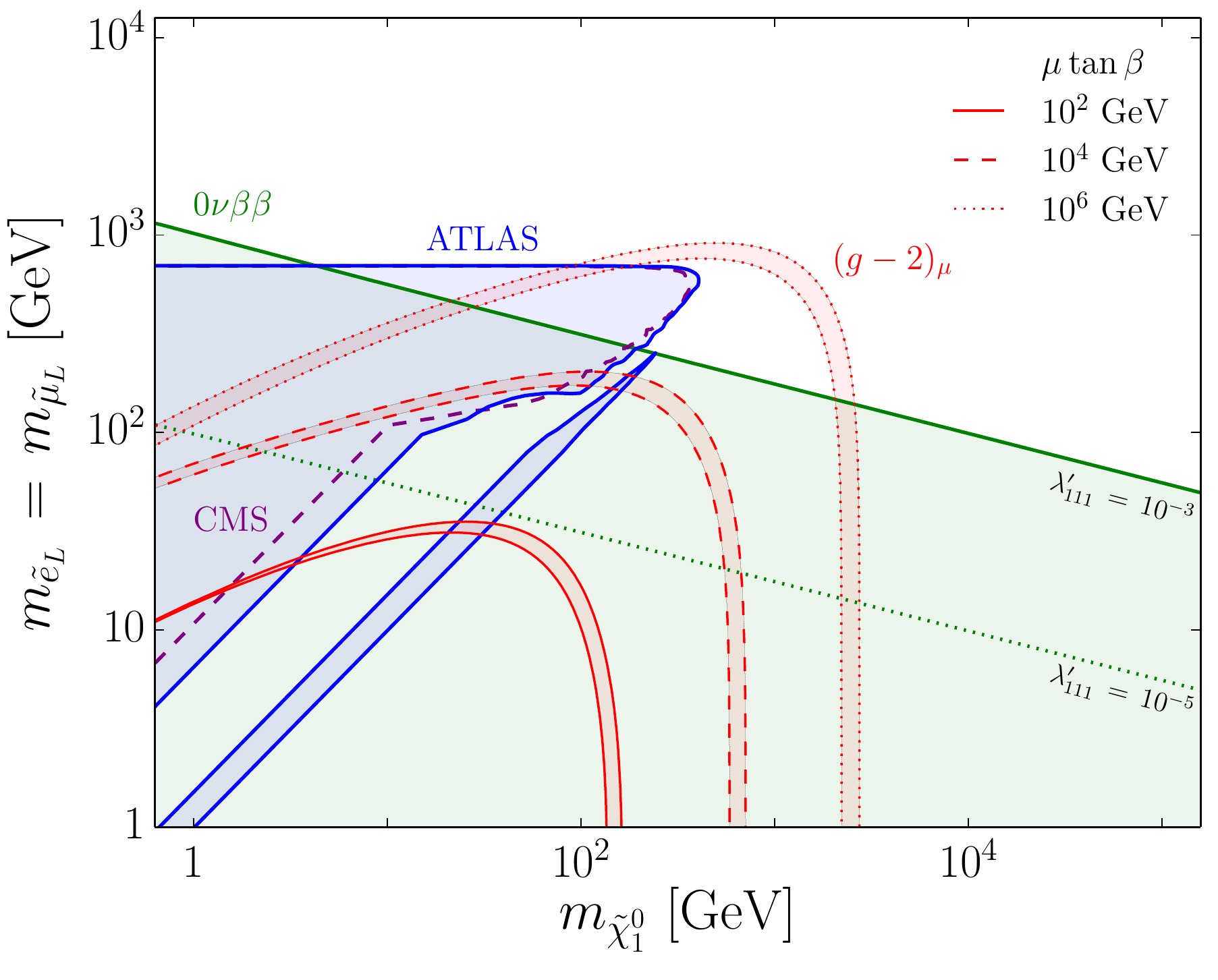}
	\caption{The constraints from $0\nu\beta\beta$ decay and ATLAS and CMS in the $(\lambda^{\prime}_{111},\, m_{\tilde{\chi}^0_1})$ plane compared to the favoured regions suggested by the $(g-2)_{\mu}$ anomaly for different values of the MSSM parameter combination $\mu\tan\beta$.}
	\label{fig:g-2_constraints}
\end{figure}

Finally, we examine the persistence of the discrepancy between the theoretical and observed muon anomalous magnetic moment $a_{\mu}$, first measured at BNL~\cite{Muong-2:2006rrc} and recently by the Muon g-2 experiment at Fermilab~\cite{Muong-2:2021ojo}. A neutralino, left-handed smuon ($\tilde{\mu}_L$) and right-handed smuon ($\tilde{\mu}_R$) will contribute to $a_{\mu}$ at the one-loop level. The size of this contribution is well-known and is summarised in Appendix \ref{sec:g-2}. Given the observed discrepancy between the theoretical~\cite{Aoyama:2020ynm} and experimental~\cite{Muong-2:2021ojo} results,  
$\Delta a_{\mu}=(2.51 \pm 0.59) \times 10^{-9}$,  we can equate this value with the contribution from the lightest neutralino and smuons, $a^{\tilde{\chi}^0_1}_{\mu}$, to draw a favoured region in the $(m_{\tilde{\mu}_L},\, m_{\tilde{\chi}^0_1})$ plane. We will also assume the left-handed selectron and smuon to be degenerate, $m_{\tilde{e}_L} = m_{\tilde{\mu}_L}$, one of the conditions of the pMSSM. In Fig.~\ref{fig:g-2_constraints}, we therefore show the favoured region from $(g-2)_{\mu}$ in the $(m_{\tilde{e}_L},\, m_{\tilde{\chi}^0_1})$ plane; the upper and lower red lines are derived by using the discrepancy plus or minus the associated $1\sigma$ uncertainty. The location of the favoured region is controlled by the product of MSSM parameters $\mu$ and $\tan\beta$ defined in Section~\ref{sec:MSSMconventions}. The $(g-2)_{\mu}$ favoured region is plotted for three values of $\mu\tan\beta$. The plot allows us to compare the $(g-2)_{\mu}$ favoured region with the collider and $0\nu\beta\beta$ decay constraints. It can be seen that ATLAS and CMS~\cite{Aad:2019vnb,Sirunyan:2020eab} exclude the favoured region from $(g-2)_{\mu}$ for $m_{\tilde{\chi}^0_1}\lesssim 100$~GeV. The $0\nu\beta\beta$ decay bound depends on $\lambda_{111}'$, but as the $(g-2)_{\mu}$ favoured region falls off very quickly for  $m_{\tilde{\chi}^0_1}\gtrsim 100$~GeV, $0\nu\beta\beta$ decay excludes this section of parameter space even for relatively small values of $\lambda_{111}'$. 

\section{Conclusions}
\label{sec:ch6-concl}

The non-observation of signs of supersymmetry at the LHC (and other searches) puts stringent constraints on this theoretical framework. It generally excludes supersymmetric particles with masses lighter than the electroweak to TeV scale. In the context of R-parity violating supersymmetry this also puts limits on the amount of observable lepton number violation in low energy experiments such as $0\nu\beta\beta$ decay. This process can be induced in R-parity violating supersymmetry by the exchange of a neutralino, which, in this context, is generally taken to be heavy. Neutralinos, on the other hand, are relatively weakly constrained if model-specific assumptions are removed, and specifically the lightest neutralino can be essentially massless within the full MSSM parameter space.

We have performed a detailed analysis of the contribution of the R-parity violating parameter $\lambda_{111}'$ to $0\nu\beta\beta$ decay; in particular, extending the possible mass of the lightest neutralino $\tilde{\chi}^0_1$ to below the Fermi momentum scale $p_{\text{F}}$ relevant to $0\nu\beta\beta$ decay, thus changing the process from short-range to long-range behaviour. In this context, we examined the current experimental constraints on the superpartners relevant to $0\nu\beta\beta$ decay; the lightest neutralino $\tilde{\chi}^0_1$, the selectron $\tilde{e}_L$, up squark $\tilde{u}_L$, down squark $\tilde{d}_R$ and gluino $\tilde{g}$. Depending on the value of the neutralino mass, collider experiments ATLAS and CMS set lower bounds on $m_{\tilde{e}_L}$, $m_{\tilde{u}_L}$, $m_{\tilde{d}_R}$ and $m_{\tilde{g}}$. However, for $m_{\tilde{\chi}^0_1} > m_{\tilde{X}}$, the ATLAS and CMS constraints vanish. While the OPAL experiment sets lower bounds in this region from a search for direct R-parity violating decays, we instead use the approximate lower limits on the superpartner masses from the pMSSM analysis of ATLAS. Direct constraints on $\lambda_{111}'$ can be set from measurements of the $V_{ud}$ element of the CKM matrix and tests of lepton universality in pion decays. Especially relevant for our analysis, we discuss the bound on $m_{\tilde{\chi}^0_1}$ and $\lambda_{111}'$ from the R-parity violating decays of the lightest neutralino and its impact on big bang nucleosynthesis. This requires the total lifetime to be $\tau_{\tilde{\chi}^0_1}< 1$~s.

Using recent $0\nu\beta\beta$ decay nuclear matrix elements, we reassess current limits and future sensitivity on the R-parity violating coupling $\lambda_{111}'$ from contributions to the $0\nu\beta\beta$ decay half-life induced by the exchange of neutralinos and gluinos. Here we specifically focus on scenarios where the neutralino can be light, i.e. close to or even below the relevant Fermi scale $p_{\text{F}}\sim 100$~MeV of $0\nu\beta\beta$ decay. We find that current limits can exclude $\lambda_{111}'$ down to $\lambda_{111}' \lesssim 10^{-3}$ for $m_{\tilde{\chi}_1^0} \approx 100$~MeV with other sparticle masses satisfying current collider constraints, and future $0\nu\beta\beta$ decays searches can improve this to $\lambda_{111}' \lesssim 10^{-4}$. While light $\tilde{\chi}_1^0$ masses, $m_{\tilde{\chi}_1^0} \lesssim 0.1 - 1$~GeV, are disfavoured in R-parity violating supersymmetry due to the impact of $\tilde{\chi}_1^0$ decays on BBN, there is a sizeable dependence on other MSSM parameters.

This demonstrates that $0\nu\beta\beta$ decay does provide important constraints on R-parity violating supersymmetry for lightest neutralino masses $m_{\tilde{\chi}_1^0} \gtrsim 10$~MeV where novel long-range contributions are induced. We note that such a scenario may have interesting consequences for the anomalous magnetic moment of the muon, namely the observed anomaly can be related to an observable $0\nu\beta\beta$ decay rate that is detectable in future searches.

\section*{Acknowledgments}

P.~D.~B. and F.~F.~D. acknowledge support from the UK Science and Technology Facilities Council (STFC) via the Consolidated Grants ST/P00072X/1 and ST/T000880/1. P.~D.~B. has received support from the European Union's Horizon 2020 research and innovation programme under the Marie Sk\l{}odowska-Curie grant agreement No 860881-HIDDeN. The work of B.~D.\ is supported in part by the US Department of Energy under Grant No.~DE-SC0017987.

\appendix

\section{Neutralino Decay Channels and Total Width}
\label{sec:decay_rates}

In this appendix we list the partial decay rates for the massive neutralinos $\tilde{\chi}^0_i$ assuming a non-zero RPV coupling $\lambda_{ijk}'$. As we are considering in this work the impact of $\lambda_{111}'$ on $0\nu\beta\beta$ decay, we limit the list to the decay channels induced by $\lambda_{111}'$ (and thus involving electron neutrinos, electrons, up quarks and down quarks). For non-zero $\lambda_{111}'$, $\tilde{\chi}^0_i$ can decay by first coupling to a first generation fermion and its superpartner via the interactions in Eq.~\eqref{eq:LRint}. If the superpartner is heavy and therefore virtual, it will decay via a $\lambda_{111}'$ interaction in Eq.~\eqref{eq:L_LQD} to two additional first generation fermions. As there will always be two quarks in the final state, these will combine to form a meson if the neutralino mass $m_{\tilde{\chi}^0_i}$ is greater than the meson mass.

To calculate the total neutralino decay rate $\Gamma_{\tilde{\chi}^0_i}$ for a given neutralino mass $m_{\tilde{\chi}^0_i}$, we use the \textit{channel-by-channel} approach for neutralino masses $m_{\tilde{\chi}^0_i}\lesssim 800$~MeV by summing the partial rates of the semi-leptonic decays $\tilde{\chi}^0_i\rightarrow \ell(\nu)\mathcal{H}$. For masses $m_{\tilde{\chi}^0_i}\gtrsim 800$~MeV, we take into account all additional semi-leptonic decay modes by using the \textit{inclusive} approach, suggested in Ref.~\cite{Gribanov:2001vv}. This approximates the sum over all the semi-leptonic decay channels to be equal to the quark-level process $\tilde{\chi}^0_i\rightarrow \ell(\nu)q_1\bar{q}_2$. While this leading-order approximation neglects perturbative and non-perturbative QCD corrections, it is expected to be accurate for $m_{\tilde{\chi}_i^0}\gg 200$~MeV~\cite{Helo:2010cw}.

The partial rates for the semi-leptonic decays, normalised to the Fermi constant squared $G_{\text{F}}^2$, are given (neglecting sfermion mixing) by
\begin{align}
\label{eq:semilep-1}
\Gamma(\tilde{\chi}^0_i \rightarrow \nu P^0) & = \frac{G_{\text{F}}^2 \eta_{\nu P}}{128\pi} m^3_{\tilde{\chi}^0_{i}}f_{P}^2 \,\frac{y^4_{P}(1-y_{P}^2)^2}{(y_u+y_d)^2}\,,\\
\Gamma(\tilde{\chi}^0_i \rightarrow  e^{-}P^{+}) & = \frac{G_{\text{F}}^2\eta_{\ell P}}{128\pi} m^3_{\tilde{\chi}^0_{i}}f_{P}^2\,\frac{y^4_{P}\lambda^{\frac{1}{2}}(y_{e}^2,y_{P}^2)(1+y_e^2-y^2_P)}{(y_u+y_d)^2}\,,\\
\Gamma(\tilde{\chi}^0_i \rightarrow \nu V^{0}) & \simeq \frac{G_{\text{F}}^2\eta_{\nu V}}{2\pi} m^3_{\tilde{\chi}^0_{i}}f_{V}^2\,(1-y_{V}^2)\big[2-y_{V}^2(1+y^2_V)\big]\,,\\
\Gamma(\tilde{\chi}^0_i \rightarrow e^{-}V^{+}) & \simeq \frac{G_{\text{F}}^2\eta_{\ell V}}{2\pi} m^3_{\tilde{\chi}^0_{i}}f_{V}^2\,\lambda^{\frac{1}{2}}(y_{e}^2,y_{V}^2)\big[2(1-y_e^2)^2-y^2_V(1+y^2_e+y^2_V)\big]\,,
\label{eq:semilep-4}
\end{align}
where $y_X = m_X / m_{\tilde{\chi}^0_{i}}$ and $P^0=\{\pi^0,\eta\}$, $P^+=\{\pi^+\}$, $V^0=\{\rho,\omega\}$ and $V^+=\{\rho^{+}\}$ are the relevant light scalar and vector mesons respectively. Here, $f_X$ are the meson decay constants, the kinematic function $\lambda(x,y)=(1-x-y)^2-4xy$,
and the dimensionless parameters $\eta_{\ell P}$ and $\eta_{\ell V}$ are given by
\begin{align}
\eta_{\ell P} &= \frac{8\pi\alpha_2|\lambda^{\prime }_{111}|^2}{G_{\text{F}}^2}\bigg|\frac{V^{LL}_{e\tilde{\chi}^0_{i}}}{m^2_{\tilde{e}_L}}-\frac{V^{LL}_{u\tilde{\chi}^0_{i}}}{2m^2_{\tilde{u}_L}}-\frac{V^{RR*}_{d\tilde{\chi}^0_{i}}}{2m^2_{\tilde{d}_R}}\bigg|^2\,,\\
\eta_{\ell V} &= \frac{\pi\alpha_2|\lambda^{\prime }_{111}|^2}{2G_{\text{F}}^2}\bigg|\frac{V^{LL}_{u\tilde{\chi}^0_{i}}}{m^2_{\tilde{u}_L}}-\frac{V^{RR*}_{d\tilde{\chi}^0_{i}}}{m^2_{\tilde{d}_R}}\bigg|^2\,.
\end{align}
The parameters $\eta_{\nu P}$ and $\eta_{\nu V}$ are found by replacing $m_{\tilde{e}_L}\rightarrow m_{\tilde{\nu}_L}$, $m_{\tilde{u}_L}\rightarrow m_{\tilde{d}_L}$, $V^{LL}_{e\tilde{\chi}^0_{i}}\rightarrow V^{LL}_{\nu\tilde{\chi}^0_{i}}$ and $V^{LL}_{u\tilde{\chi}^0_{i}}\rightarrow V^{LL}_{d\tilde{\chi}^0_{i}}$ in $\eta_{\ell P}$ and $\eta_{\ell V}$ respectively. The rates for the processes $\tilde{\chi}^0_i \rightarrow \bar{\nu} P^0$, $\tilde{\chi}^0_i \rightarrow e^+ P^-$, $\tilde{\chi}^0_i \rightarrow \bar{\nu} V^0$ and $\tilde{\chi}^0_i \rightarrow e^{+} V^-$ are also given by Eqs.~\eqref{eq:semilep-1} to \eqref{eq:semilep-4} respectively.

Note that we have only included mesons with masses up to $m_\omega = 782.65$~GeV in Eqs.~\eqref{eq:semilep-1} to \eqref{eq:semilep-4}. Above this mass the uncertainty on the appropriate decay constants increases considerably and we instead compute the total (semi-leptonic) neutralino decay rate above the scale $\mu_0\sim 1$~GeV using the inclusive approach. The three-body decays to a lepton and two quarks are found to be
\begin{align}
\Gamma(\tilde{\chi}^0_i \rightarrow \nu d\bar{d}) & = \frac{G_{\text{F}}^2\eta_{\nu dd}}{1024\pi^3}m_{\tilde{\chi}^0_{i}}^5\,,\\
\Gamma(\tilde{\chi}^0_i \rightarrow e^{-}u\bar{d}) & = \frac{G_{\text{F}}^2\eta_{\ell ud}}{1024\pi^3}m_{\tilde{\chi}^0_{i}}^5\,,
\end{align}
where $\eta_{\ell ud}$ is given by
\begin{align}
\eta_{\ell ud} &= \frac{\pi\alpha_2|\lambda^{\prime }_{111}|^2}{G_{\text{F}}^2}\left[\frac{4\big|V^{LL}_{e\tilde{\chi}^0_{i}}\big|^2}{m^4_{\tilde{e}_L}}+\frac{241\big|V^{LL}_{u\tilde{\chi}^0_{i}}\big|^2}{m^4_{\tilde{u}_L}}+\frac{241\big|V^{RR}_{d\tilde{\chi}^0_{i}}\big|^2}{m^4_{\tilde{d}_R}}\right.\nonumber\\
&\left.\hspace{7em}-\frac{4\,\text{Re}\big[V^{LL}_{e\tilde{\chi}^0_{i}}V^{LL*}_{u\tilde{\chi}^0_{i}}\big]}{m^2_{\tilde{e}_L}m^2_{\tilde{u}_L}}-\frac{4\,\text{Re}\big[V^{LL}_{e\tilde{\chi}^0_{i}}V^{RR}_{d\tilde{\chi}^0_{i}}\big]}{m^2_{\tilde{e}_L}m^2_{\tilde{d}_R}}-\frac{478\,\text{Re}\big[V^{LL}_{u\tilde{\chi}^0_{i}}V^{RR}_{d\tilde{\chi}^0_{i}}\big]}{m^2_{\tilde{u}_L}m^2_{\tilde{d}_R}} \right]\,.
\end{align}
The parameter $\eta_{\nu dd}$ is again found by replacing $m_{\tilde{e}_L}\rightarrow m_{\tilde{\nu}_L}$, $m_{\tilde{u}_L}\rightarrow m_{\tilde{d}_L}$, $V^{LL}_{e\tilde{\chi}^0_{i}}\rightarrow V^{LL}_{\nu\tilde{\chi}^0_{i}}$ and $V^{LL}_{u\tilde{\chi}^0_{i}}\rightarrow V^{LL}_{d\tilde{\chi}^0_{i}}$ in $\eta_{\ell ud}$. The total (semi-leptonic) decay width is therefore
\begin{align}
\Gamma(\tilde{\chi}^0_i \rightarrow  e^{-}(\nu)\mathcal{H}) &=
\Theta(\mu_0-m_{\tilde{\chi}_i^0})\sum\big[\Gamma(\tilde{\chi}^0_i \rightarrow \nu P^0)+\Gamma(\tilde{\chi}^0_i \rightarrow e^{-} P^+)\nonumber\\
&\hspace{9em}+\Gamma(\tilde{\chi}^0_i \rightarrow \nu V^0)+\Gamma(\tilde{\chi}^0_i \rightarrow e^{-} V^+)\big]\nonumber\\ &\quad~+\Theta(m_{\tilde{\chi}_i^0}-\mu_0)\big[\Gamma(\tilde{\chi}^0_i \rightarrow \nu d\bar{d}) + \Gamma(\tilde{\chi}^0_i \rightarrow e^{-}u\bar{d})\big]\,,
\end{align}
where the sum is over pseudoscalar and vector mesons lighter than $m_{\omega}$.

For neutralino masses below the neutral pion mass, $m_{\tilde{\chi}^0_i} < m_{\pi^0}$, the decays above are no longer kinematically accessible. Instead, light neutralinos may decay radiatively to a photon and neutrino via a loop containing a down quark and a left- or right-handed down squark. The rate for this process is approximately
\begin{align}
\Gamma(\tilde{\chi}^0_i \rightarrow \nu\gamma)\sim \frac{\alpha G_{\text{F}}}{768\pi^3}\frac{\alpha_2|\lambda'_{111}|^2}{ G_{\text{F}}}m^3_{\tilde{\chi}^0_i}m_d^2\left[\frac{\big|V^{LL}_{d\tilde{\chi}^0_i}\big|^2}{m_{\tilde{d}_L}^4}F\Bigg(\frac{m_d^2}{m_{\tilde{d}_L}^2}\Bigg)+\frac{\big|V^{RR}_{d\tilde{\chi}^0_i}\big|^2}{m_{\tilde{d}_R}^4}F\Bigg(\frac{m_d^2}{m_{\tilde{d}_R}^2}\Bigg)\right]\,,
\end{align}
with $F(x) = \left(\ln(x)+3/2\right)^2$~\cite{Hall:1983id,Dawson:1985vr}.
Finally, light neutralinos can also decay to a gravitino $\tilde{\psi}$ and photon if $m_{\tilde{\psi}}<m_{\tilde\chi}^{0_i}$, giving the partial decay rate
\begin{align}
\Gamma(\tilde{\chi}^0_i \rightarrow \tilde{\psi} \gamma)=\frac{c_{W}^{2}}{48 \pi m_{P}^{2}} \frac{1}{m^3_{\tilde{\chi}^0_{i}}m_{\tilde{\psi}}^{2}}\big(m_{\tilde{\chi}^0_{i}}^{2}-m_{\tilde{\psi}}^{2}\big)^{3}\big(m_{\tilde{\chi}^0_{i}}^{2}+3 m_{\tilde{\psi}}^{2}\big)\,,
\end{align}
where $m_P = 1.22\times 10^{22}$~GeV is the Planck mass~\cite{Covi:2009bk}. 

The total neutralino decay rate is thus given by
\begin{align}
\label{eq:total_width}
\Gamma_{\tilde{\chi}^0_i} = \Gamma(\tilde{\chi}^0_i \rightarrow \tilde{\psi} \gamma) + 2\Gamma(\tilde{\chi}^0_i \rightarrow \nu \gamma) + 2\Gamma(\tilde{\chi}^0_i \rightarrow  e^{-}(\nu)\mathcal{H}),
\end{align}
where we have multiplied the partial decay rates for $\tilde{\chi}^0_i \rightarrow \nu \gamma$ and $\tilde{\chi}^0_i \rightarrow  e^{-}(\nu)\mathcal{H}$ by a factor of two to account for decays to charge conjugate final states. We use Eq.~\eqref{eq:total_width} to exclude regions of the parameter space by enforcing the naive bound from BBN, $\tau_{\tilde{\chi}^0_1} = \Gamma_{\tilde{\chi}^0_1}^{-1} \lesssim 1$~s.

\section{Muon Anomalous Magnetic Moment}
\label{sec:g-2}

The one-loop contribution of the neutralinos and smuons to the muon anomalous magnetic moment is well-known and is given by
\begin{align}
\label{eq:amu}
a_{\mu}^{\tilde{\chi}^{0}}=-\sum_{i, \kappa} \frac{\alpha_2}{2 \pi} \frac{m_{\mu}^{2}}{m_{\tilde{\ell}_{\kappa}}^{2}}\left[\frac{1}{12} \mathcal{A}_{\tilde{\chi}^0_{i}\tilde{\ell}_\kappa} F_{1}\Bigg(\frac{m_{\tilde{\chi}^0_{i}}^2}{m_{\tilde{\ell}_\kappa}^2}\Bigg)+\frac{m_{\tilde{\chi}^0_{i}}}{6 m_{\mu}} \mathcal{B}_{\tilde{\chi}^0_{i}\tilde{\ell}_\kappa} F_{2}\Bigg(\frac{m_{\tilde{\chi}^0_{i}}^2}{m_{\tilde{\ell}_\kappa}^2}\Bigg)\right]\,,
\end{align}
where the indices $i$ and $\kappa$ label the neutralino and slepton mass eigenstates respectively~\cite{Grifols:1982vx,Moroi:1995yh,Stockinger:2006zn,Fargnoli:2013zda,Fargnoli:2013zia}. The loop functions are given by
\begin{align}
F_{1}(x) & = \frac{2}{(1-x)^4}\big[(1-x)(1-5x-2x^2)-6x^2\ln x\big] \,,\\
F_{2}(x) & = \frac{3}{(1-x)^3}\big[(1-x)(1+x)+2x\ln x\big] \,,
\end{align}
while the factors $\mathcal{A}_{\tilde{\chi}^0_{i}\tilde{\ell}_\kappa}$ and $\mathcal{B}_{\tilde{\chi}^0_{i}\tilde{\ell}_\kappa}$ are
\begin{align}
\label{eq:AandB}
\begin{aligned}
\mathcal{A}_{\tilde{\chi}^0_{i}\tilde{\ell}_\kappa} & \equiv \big|(V^L_{\mu\tilde{\chi}^0_{i}})_{2\kappa}\big|^2 + \big|(V^R_{\mu\tilde{\chi}^0_{i}})_{2\kappa}\big|^2\,, \\
\mathcal{B}_{\tilde{\chi}^0_{i}\tilde{\ell}_\kappa} & \equiv (V^L_{\mu\tilde{\chi}^0_{i}})_{2\kappa} (V^R_{\mu\tilde{\chi}^0_{i}})^*_{2\kappa} + (V^R_{\mu\tilde{\chi}^0_{i}})_{2\kappa} (V^L_{\mu\tilde{\chi}^0_{i}})^*_{2\kappa}\,,
\end{aligned}
\end{align}
where the left- and right-handed neutralino couplings are given in Eqs.~\eqref{eq:LRcouplings} and \eqref{eq:LRcouplings2}. 

In this work we neglect for simplicity the generational mixing of charged sleptons. For the exchange of the lightest neutralino in the one-loop diagram, we therefore need the coefficients $\mathcal{A}_{\tilde{\chi}^0_{1}\tilde{\ell}_2}$, $\mathcal{A}_{\tilde{\chi}^0_{1}\tilde{\ell}_5}$, $\mathcal{B}_{\tilde{\chi}^0_{1}\tilde{\ell}_2}$ and $\mathcal{B}_{\tilde{\chi}^0_{1}\tilde{\ell}_5}$, where $\tilde{\ell}_2 \equiv \tilde{\mu}_1$ and $\tilde{\ell}_5\equiv \tilde{\mu}_2$ are the smuon mass eigenstates. In the left- and right-handed basis for the smuons, the $2\times2$ mass matrix squared is
\begin{align}
\label{eq:smuonmatrix}
\mathcal{M}^{2}_{\tilde{\mu}}=
\begin{pmatrix}
M^2_{\tilde{\mu}LL} & M^2_{\tilde{\mu}RL}  \\
M^2_{\tilde{\mu}LR} & M^2_{\tilde{\mu}RR} \\
\end{pmatrix}\,,
\end{align}
where,
\begin{align}
M^2_{\tilde{\mu}LL} &= m_{\tilde{L}}^2 + m_{\mu}^2 + m_{Z}^2\left(-\frac{1}{2}+s_W^2\right)c_{2\beta}\,, \\
M^2_{\tilde{\mu}RR} &= m_{\tilde{E}}^2 + m_{\mu}^2 - m_{Z}^2s_W^2c_{2\beta}\,,\\
M^2_{\tilde{\mu}RL} &= M^{2\dagger}_{\tilde{\mu}LR} = A_{\tilde{E}} v c_{\beta} - m_{\mu}\mu^*t_{\beta} \,.
\end{align}
Here, $m_{\tilde{L}}$, $m_{\tilde{E}}$ are the standard left- and right-handed smuon masses, $A_{\tilde{E}}$ is the smuon $A$-term, and the other parameters are defined in the main text. The matrix in Eq.~\eqref{eq:smuonmatrix} can be diagonalised by a $2\times 2$ unitary matrix to give the following squared masses for the mass eigenstate smuons $\tilde{\mu}_1$ and $\tilde{\mu}_2$,
\begin{align}
m^2_{\tilde{\mu}_{1,2}} &= \frac{1}{2}\left[\text{Tr}\mathcal{M}^2_{\tilde{\mu}}\mp \sqrt{(\text{Tr}\mathcal{M}^2_{\tilde{\mu}})^2-4\,\text{det}\mathcal{M}^2_{\tilde{\mu}}}\right] \,.
\end{align}
The mixing matrix contains the mixing angle $\theta_{\tilde{\mu}}$, given by
\begin{align}
\label{eq:smuonmixing}
\sin 2\theta_{\tilde{\mu}} = \frac{2|A_{\tilde{E}} v c_{\beta}-m_{\mu}\mu t_{\beta}|}{m^2_{\tilde{\mu}_2}-m^2_{\tilde{\mu}_1}}\,,
\end{align}
where we have assumed the parameters $A_{\tilde{E}}$ and $\mu$ to be real. Considering the exchange of the lightest neutralino (with $N_{i1} = 1$ and $N_{i2} = N_{i3} = N_{i4} = 0$) and the two mass eigenstate smuons, the coefficients in Eq.~\eqref{eq:AandB} are
\begin{align}
\begin{aligned}
\mathcal{A}_{\tilde{\chi}^0_1\tilde{\mu}_1} & = \frac{t_W^2}{4}(1+3\sin^2\theta_{\tilde{\mu}})\,,\\
\mathcal{A}_{\tilde{\chi}^0_1\tilde{\mu}_2}  &= \frac{t_W^2}{4}(1+3\cos^2\theta_{\tilde{\mu}})\,,\\
\mathcal{B}_{\tilde{\chi}^0_1\tilde{\mu}_1} & = -\mathcal{B}_{\tilde{\chi}^0_1\tilde{\mu}_2} = -\frac{t_W^2}{2}\sin2\theta_{\tilde{\mu}}\,.
\end{aligned}
\end{align}
It is now common in the literature to take the limit of degenerate smuon mass eigenstates, $m_{\tilde{\mu}_1} = m_{\tilde{\mu}_2}$. However, in this work we are interested in a wider range of parameter space (such as large values of $\mu \tan\beta$) where this limit is no longer applicable. To calculate the anomalous magnetic moment, we therefore rearrange Eq.~\eqref{eq:smuonmixing} for $m_{\tilde{\mu}_2}$ and insert into Eq.~\eqref{eq:amu}, setting $\theta_{\tilde{\mu}}=\frac{\pi}{4}$.

\bibliographystyle{JHEP}
\bibliography{references-4}
\end{document}